\documentclass[journal]{IEEEtran}

\usepackage{multirow}
\usepackage{ifpdf}
\usepackage{float}
\usepackage[pdftex]{graphicx}
\usepackage{import}
\usepackage{rotating}
\usepackage{array}
\usepackage{stfloats}
\usepackage{wrapfig}
\usepackage{cite}

\begin{document}

\title{Integrated Photonic Tensor Processing Unit for a Matrix Multiply: a Review}

\author{Nicola Peserico, Bhavin J. Shastri,
        ~\IEEEmembership{Senior Member,~IEEE,} Volker J. Sorger*,~\IEEEmembership{Fellow, Optica, Senior Member, IEEE}
\thanks{N. Peserico, V.J. Sorger are with the Department of Electrical and Computer Engineering, George Washington University. *E-mail: sorger@gwu.edu (see sorger.seas.gwu.edu)}
\thanks{V.J. Sorger is with Optelligence LLC}
\thanks{B. J. Shastri is with the Department of Physics, Engineering Physics \& Astronomy, Queen's University, Kingston, ON K7L 3N6, Canada.}
\thanks{Manuscript received xxx ; revised yyy.}}


\maketitle

\begin{abstract}
The explosion of artificial intelligence and machine-learning algorithms, connected to the exponential growth of the exchanged data, is driving a search for novel application-specific hardware accelerators. Among the many, the photonics field appears to be in the perfect spotlight for this global data explosion, thanks to its almost infinite bandwidth capacity associated with limited energy consumption. In this review, we will overview the major advantages that photonics has over electronics for hardware accelerators, followed by a comparison between the major architectures implemented on Photonics Integrated Circuits (PIC) for both the linear and nonlinear parts of Neural Networks. By the end, we will highlight the main driving forces for the next generation of photonic accelerators, as well as the main limits that must be overcome.
\end{abstract}


\IEEEpeerreviewmaketitle
\section{Introduction}
\begin{figure*}[t!]
    \centering
    \includegraphics[width=0.7\textwidth]{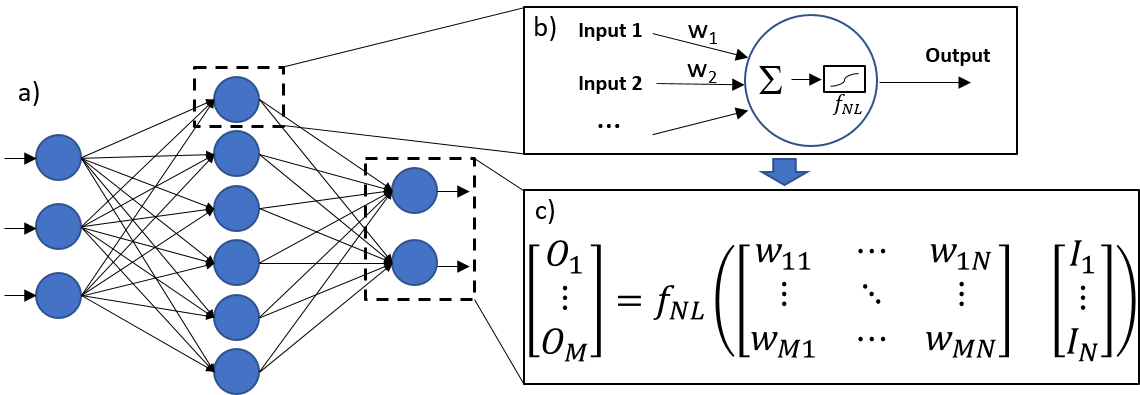}
    \caption{Breaking down of a Fully-Connected Neural Network. (a) Example of a NN having one hidden layer. (b) Every single neuron receives the input signals from all the previous layer neurons, scaled by a factor $w$, performing their summation and passing through an activation function. (c) This single neuron can be generalized by including the whole layer, employing a matrix representation. }
    \label{fig:NN}
\end{figure*}

\IEEEPARstart{T}{he} latest decade has seen the exponential growth of Machine Learning (ML) as one of the main branches of the Artificial Intelligence field\cite{sarker2021machine}. At the core of this branch, there is the assumption that a machine can learn to perform any task if a training algorithm is applied.  While historically the concept of ML can be tracked back from the '50s\cite{5392560,fradkov2020early}, just in recent decades the concept has started to attract more and more interest\cite{yadav2015history}, thanks to the improvement of the mathematical approaches (such as back-propagation\cite{Leonard1990}), and computation capabilities, that allowed to run complex ML algorithms.\\
To implement ML applications, several algorithms and circuits have been proposed\cite{mahesh2020machine}. One approach relies on mimicking the human brain structure, which has led to several implementations, where Neural Networks (NNs) have become the most popular (fig. \ref{fig:NN}), thanks to its flexibility and scalability\cite{schmidhuber2015deep, abiodun2018state}. A NN is formed by a sequence of interconnected layers of neurons, whose inputs are the output of all the neurons of the previous layer (fig. \ref{fig:NN}a). The output of a single neuron is the result of the scaled linear summation of the input passed by an activation (nonlinear) function (fig. \ref{fig:NN}b). In this framework, a whole layer can be seen as matrix multiplication, followed by the activation function, allowing for a more straightforward implementation on hardware. The values used to scale the inputs (the $W$ matrix) are the learning parameters that the NN needs to compute using the selected method (i.e. back-propagation). By so, for each NN, we can see two separate steps: the training one, where all the parameters are computed using training algorithms and dataset, and the second one, called inference or classification, where the NN is used over a novel set of data input. Research on NN has brought other implementations for each layer, based on the application and/or input. For example, convolution layers are widely used in the image and video context, where a certain trainable filter is applied to a portion of a 2D input\cite{rawat2017deep}. More and more complex tasks can be performed by NN by adding more and more layers implementing Deep Neural Networks (DNNs) for Deep Learning.\\
After the initial creation of the ML concept, followed by a winter phase due to the lack of hardware\cite{fradkov2020early}, ML has raised again following the exponential increase of computer performance, creating an environment where DNN can have tens of layers and millions of parameters. One example that has shown all the potential of this approach is called DALL$\cdot$E2, one of the most advanced text-to-image DNN, with over 3.5 billion parameters\cite{ramesh2022hierarchical}.\\
Such large and extended networks raise an enormous demand in terms of computational power\cite{hwang2018computational}, challenging current hardware technologies in terms of operation per second, latency, and power consumption. The flexibility and scalability of digital electronics have allowed the creation of a framework where NNs can be coded, tested, and used\cite{erickson2017toolkits}. As the NN became larger and larger, the digital approach started to look for novel solutions to keep pace and deliver enough performance levels to run the NN\cite{reuther2019survey}. Those solutions are based on scaling, by using interconnected hardware in data centers, or by architecture changes, for example moving from generic CPU to application- or numerical- specific ones, such as FPGA, GPU, or ASIC, called Tensor Core\cite{suda2016throughput, zhou2017tunao, reuther2020survey}. However, some of the limitations still exist, due to more physical reasons, such as energy consumption and latency\cite{han2015learning}. For these reasons, research has started to look for novel technologies that can provide a better hardware accelerator for NNs. Optics (and photonics) have been raised as an alternative approach for hardware implementation of NN, thanks to its speed-of-light latency and low energy consumption\cite{Shastri2021,ma2021deep, Miscuglio2020, miller2017attojoule, sunny2021survey}. Moreover, Silicon Photonics has started to become a reliable and diffuse technology, allowing the implementation of Photonic Neural Network (PNN) hardware accelerator at the chip scale, to better fit the needs of final users\cite{soref1987electrooptical,chrostowski2015silicon,siew2021review}.\\
In this paper, we will review why and how silicon photonics chips have addressed the challenge of providing a hardware accelerator for PNN. After an initial part on electronics limitations and photons potential in this field, we will look into the main implementations of Photonic Tensor Core (PTC), either based on coherent interference or WDM/MDM approaches. We will address the limitations and scalability of such solutions, focusing on the most challenging part related to the activation function. We will conclude with a discussion of what the near and long-term future look like for such PNNs.\\

\section{Electronics vs. Photonics}
Digital electronics has been the hardware foundation that allowed the growth of NNs since it can provide flexibility, scalability, and fast delivery times. Even if the Von Neumann architecture is not the best one for NN applications\cite{haensch2017scaling}, it has provided the right framework to develop NNs in their early stage. 
Moreover, the diffusion of programming languages for software development, and the following NN-specific libraries, has permitted the spread of NN applications since the ’90\cite{yadav2015history}. The continuous improvement in computer performances (in terms of processors, memory, and network) thanks to the development of smaller and more dense CMOS transistors\cite{etiemble201845}, has permitted to keep pace with the increasing complexity of NNs. \\
However, in the last decade, the complexity, layer density, and datasets size have evolved to a scale that a single CPU cannot handle, neither for inference nor training\cite{petrenko2018limitations}. The main limitations come from the size of the NN, which could require millions of parameters, and so the memory size and throughput become important bottlenecks, as well as the limited capability of CPU to perform float multiplication and summation, that are required for every neural layer, as shown before. All these aspects have pushed also the energy consumption related to the NN\cite{ganguly2019towards}, for both training and inference, posing an additional challenge from the hardware perspective. \\
To overcome such limitations, several paths and solutions have been explored and adopted, from both software and hardware sides. From the software and theoretical side, several strategies and optimizations have been proposed. For example, model compression allows the reduction of the number and size of weights, and by so reducing the need to transfer them from the processing unit to the memory and vice versa\cite{cheng2017survey,deng2020model,Ding2017}. Many studies have shown how the whole system's power consumption can be easily dominated by the access cost per bit to off-chip DRAM memory\cite{han2015deep}. Some of these strategies include weight quantization\cite{sung2015resiliency}, connection pruning\cite{blalock2020state}, low rank approximation\cite{astrid2018deep}, and low bit weights\cite{zhuang2018towards}.  From the hardware side, there have been two main shifts: the first takes advantage of the computation parallelization, and the second push for more application-specific hardware, in particular on the math unit. By using multiple systems in a balanced scheme, it is possible to parallelize the layer computation over different systems, and so assure a more high throughput, even for DNN\cite{pethick2003parallelization}. Today's market presents many data centers and cloud services that provide these types of schemes, from Google Cloud to Amazon Web Service\cite{saiyeda2017cloud}. The diffusion and expansion of those data centers have reached a threshold regarding their power consumption pace rate\cite{yang2017method,rani2021survey}. The second approach works directly on the hardware optimization connected to the computation part of the NN\cite{machupalli2022review}. Since CPUs provide a limited amount of resources for math computation, NNs have moved toward GPUs, which provide faster and more specific hardware to perform float multiplication and accumulation (MAC), as a key task for each NN. The main acceleration of GPUs over CPUs is an increased number of ALU (Arithmetic Logic Unit) cores to parallelize MAC operations, roughly 1000 vs. 10, respectively. Following this trend, the use of ASIC and Tensor Processing Units (TPUs) has grown in recent years, where the actual hardware can implement the required tasks in a heavily optimized fashion as they are written in the electronic architecture\cite{suda2016throughput, zhou2017tunao, reuther2020survey}. TPUs continued the GPU push, reaching about 32,000 cores, but also added reduced memory access by deploying an systolic array, which uses an approach of featuring an array thus processing once input vector at the same time\cite{zhang2018analyzing}. Examples of ASIC can be found in many companies, such as Nvidia, Intel, and Tesla\cite{Burgess19nvidia, Yang19Intel, BannonVST19self}.\\
Even with those optimizations, digital electronic presents important limitations for NN implementations. For example, speed is always limited by the clock cycles and transistors' energy consumption, as it has been for CPUs, capping the clock to a few GHz. Moreover, the latency in the computation can be dominant, since float MAC operations still require several cycles to be performed. For applications where timing and energy consumption are a concern, such as autonomous driving for small drones, those limitations pose complex challenges to the NN engineers. \\
Optics and photonics have been raised as one of the possibilities to overcome these limitations\cite{Shastri2021,Huang2022}. The use of photons instead of electrons allows a virtually infinite bandwidth, speed-of-light propagation latency, and almost zero power consumption, thanks to the lack of RC wire charge connected to the propagation of electrons\cite{miller2017attojoule,Tait2022quantify}. Silicon Photonics, in particular, is in the right spot to provide the next generation of hardware accelerators for PNNs\cite{sunny2021survey}, thanks to the important progress that happened in the last decade\cite{hochberg2010towards,rahim2021taking,Lischke2022}, such as component density, laser integration, high-speed ($>100$ GHz) modulators and photodetectors, and low propagation losses. Other benefits that photonics has over digital MAC accelerators include 1) the ability to perform summation in the analog domain at full bit precision before ADC quantization happens; 2) temporal pooling of data such as for convolution operations by increasing the integration time of the receiver, which also lower ADC requirements; 3) high-level of fan-out via copying data passively; 4) energy-free Fourier transformation via the Fourier Theorem performing a passive FFT by an optical lens\cite{miscuglio2020massively} (i.e. also on-chip\cite{peserico2022design}); 5) the possibility to process image or lidar input directly as light signals. 
As we will see in the next section, several Photonic Integrated Circuits (PICs) have been presented in this field, showing the potential of such Photonic Tensor Cores (PTCs) in real applications. \\
It has to make clear that photonics brings its challenges too, from the energy cost of moving back and forth from the digital domain (from where data come from) to the analog (the optical) one, to the noise management for high bandwidth that limits the bit resolution at the output. Other aspects are related to the architecture implementations, as photons require an electrical system to be controlled and keep operational, making each PIC strongly related to an FPGA/ASIC that must assure its working operations\cite{cheng2020silicon,morichetti2014breakthroughs}.\\

\section{Photonics Integrated Circuit for NN: Architectures}
Several PIC architectures have been proposed over the last years to perform the Tensor Core tasks for PNNs\cite{zhou2022photonic,moss2022photonic,Alqadasi2022}. Considering the main PTC task, the MAC operation benefits from the coherent electromagnetic nature of the light, implying the possibility to perform multiplication by lossless interference, while the accumulation is performed directly on the photodector once light signals are collected. Moreover, by allowing manipulation of light employing nanoscale waveguides, PIC can integrate a large number of MAC operations on small scale, employing a high number of inputs, high-speed modulators, and photodetectors.\\ 
To perform the MAC function, several different approaches have been proposed during the latest years, varying the basic components elements, as well as the input, the weights, and the output configurations. Those different architectures show different performances, in terms of actual speed (measured as MAC operations per second), footprint, energy consumption, etc\cite{sunny2021survey}. \\ 
Here, we will review these approaches integrated into PICs, as we focus on the main differences among the architectures. Several figure-of-merits are commonly used to compare different PTC, such as MAC operation per second, or footprint\cite{Zhou2022,moss2022photonic}. They come from a system-level perspective, and are easily comparable among different architectures, even across different domains. However, for the photonics field, they mainly depend on both the technologies used for modulators (for input vectors) or the photodiodes (for output vectors) used in each implementation, which follow the possibilities given by the foundries and rarely are due to architecture choices\cite{margalit2021perspective,Bogaerts2018,Chrostowski2019}. 
Following that, it is more interesting to focus on common limitations, such as the number of controllers that each circuit requires, the footprint scaling, and the possibility to implement nonvolatile memory elements, such as Photonic RAM (P-RAM) components using Phase Change Materials (PCMs)\cite{rios2015integrated, cheng2018device}, to further reduce energy consumption. Those figure-of-merits better describe the differences between different circuits, showing that trade-offs must be addressed to evolve into this field.\\
To start the review, we first divided the PIC into two main categories, based on the mathematical approaches for the MAC operation: the first one relies on the singularization, where the main matrix is divided by the meaning of singular value decomposition into 3 matrices; the second approach avoids this decomposition, by implementing schemes that directly reflect the main matrix.\\
\subsection{$Y =  (V^T \Sigma U) X$}
\begin{figure*}[t!]
    \centering
    \includegraphics[width=\textwidth]{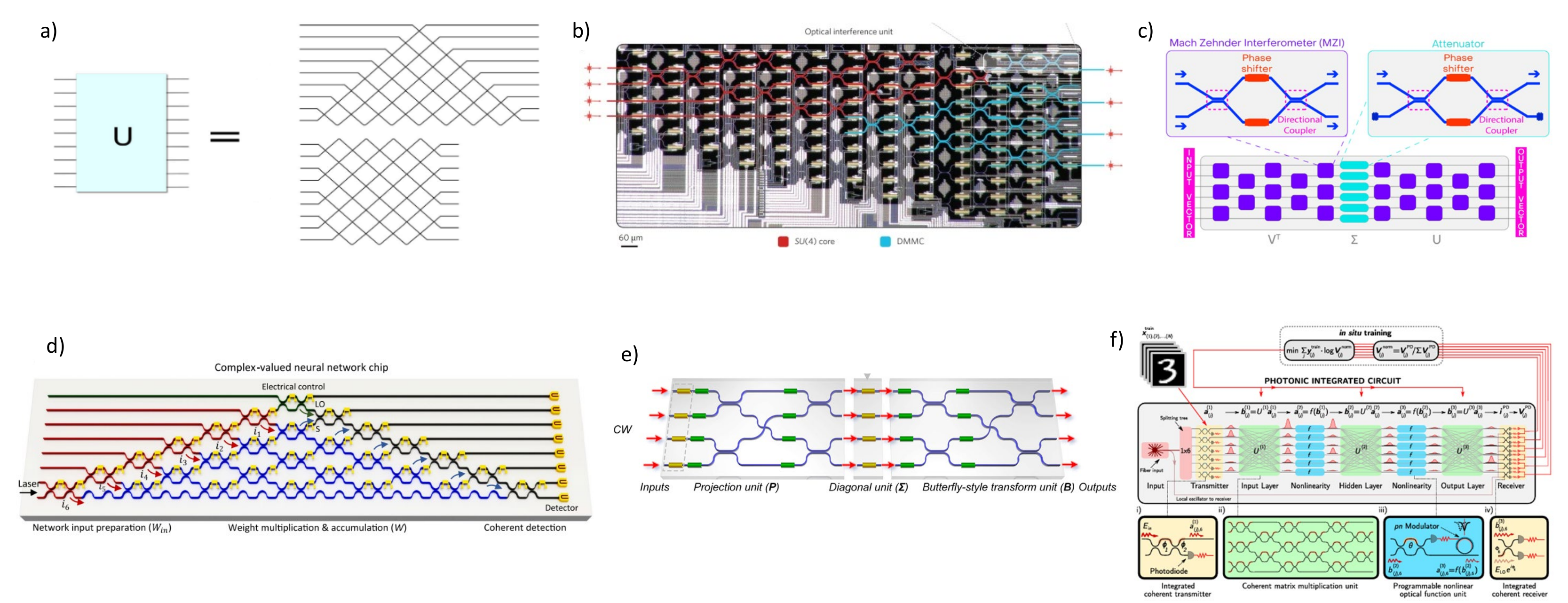}
    \caption{Examples of Photonic Integrated Circuit for Neural Networks, using Mach-Zehnder Interferometers mesh. (a) Comparison between the original Reck mesh and the improvement proposed by Clements et al.\cite{Metcalf2016}. (b) Photo of the Optical Interference Unit proposed by Shen et al.\cite{Shen2017}. (c) Similar architecture, showing the actual SVD with central attenuators line, proposed by Demirkiran et al.\cite{Demirkiran2021}. (d) Reck mesh implementing complex values for Neural Networks, done by Zhang et al.\cite{Zhang2021}. (e) Butterfly solution, exploiting pruning, presented by Feng et al.\cite{Feng2021}. (f) Full integrated Neural Network, using MZIs mesh and integrated activation functions, from Bandyopadhyay et al.\cite{Bandyopadhyay2022}  }
    \label{fig:SVD}
\end{figure*}

\begin{figure}[!]
    \centering
    \includegraphics[width=0.45\textwidth]{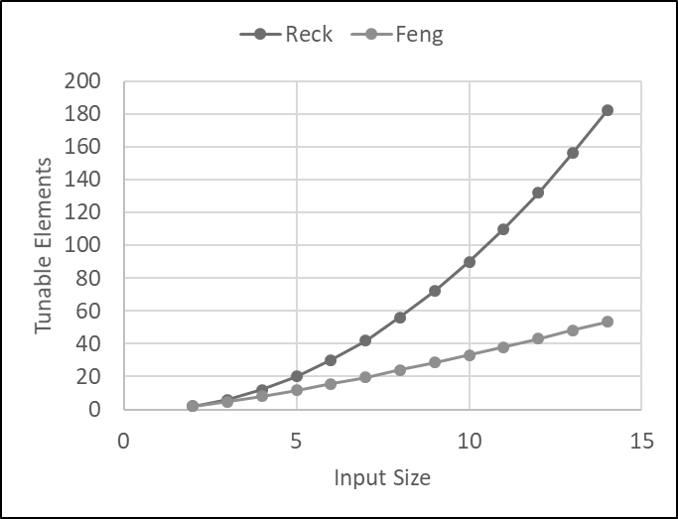}
    \caption{Scaling comparison between the original Reck scheme\cite{Reck1994}, and the one proposed by Feng et al.\cite{Feng2021}, that uses pruning to reduce the connections\cite{Ding2017}.}
    \label{fig:MZM_scal}
\end{figure}

One type of PICs exploits the single value decomposition (SVD) of matrices where the main weight matrix is divided into 3 matrices, that can be directly implemented by using cascaded Mach-Zehnder Interferometers (MZIs). This approach has its root in a work by Reck et al. in 1994\cite{Reck1994}, where they describe a simple algorithm for the realization of any $NxN$ unitary matrix. By using the SVD, the external matrices $V$ and $U$ are unitary matrices, so the implementation can be straightforward by using interconnected MZIs, while the diagonal matrix $\Sigma$ can be implemented by a series of attenuators, usually implemented by MZIs too. A more complete description and discussion were later provided by Miller et al. in 2013\cite{miller2013self}. Some examples of this architecture are shown in fig. \ref{fig:SVD} \\
The first experimental implementations were presented for quantum optics, by Carolan et al.\cite{Carolan2015}, where 15 MZI were integrated into one single silicon photonic chip. The work was followed by Riberio et al., demonstrating a $4x4$-port universal linear circuit\cite{Ruocco2016}, and by Annoni et al., presenting a mode demultiplexer with the same MZIs architecture\cite{annoni2017unscrambling}. \\
A theoretical discussion was presented by Clements et al. in 2016 on the MZI layout\cite{Metcalf2016}, shown in fig. \ref{fig:SVD}a. The paper shows a way to implement the same MZI mesh network more compactly, allowing to reduce of the insertion loss due to the shorter path length, without any mathematical limitation in the starting unitary $U$ matrix. To be noticed, this novel approach reduces the length of the device but does not reduce the number of components required. \\
The first implementation of the MZM mesh as a PTC device for NN comes from Shen et al. in 2017\cite{Shen2017} (fig. \ref{fig:SVD}b). 
The MZI mesh was used as part of an Optical Neural Network (ONN) in a Deep Learning scheme performing vowel recognition. The chip integrated 56 MZIs, showing good accuracy data and opening the path for more ONN as a way to improve energy efficiency and computational speed. 
From part of this work, a spin-off company was created and recently started to show its architecture\cite{Demirkiran2021} (fig. \ref{fig:SVD}c). In this case the silicon photonic chip has the same MZI mesh approach, but it integrates directly all the 3 matrices of the SVD, together with integrated photodetectors. The work shows 8-bit precision and the clear leverage that photonics can provide to AI accelerators in terms of energy efficiency per operation. 
A step forward was been done by Zhang et al. as they implemented a PNN with complex values, using the original Reck MZI scheme\cite{Zhang2021} (fig. \ref{fig:SVD}d). \\
While all these implementations allow having a full matrix, and so to implement a fully connected neural layer, a recent trend following the electronic approach is exploring pruning as a technique to reduce the number of connections between layers. One example in the photonic field has been presented by Feng et al.\cite{Feng2021}, shown in fig. \ref{fig:SVD}e. In this case, the matrices $V$ and $U$ are substituted with projection and transform units, that have a large reduction of the number of MZI\cite{Ding2017}. The authors show that, despite the reduction in the number of MZI, the PNN was capable to perform digit recognition over MNIST dataset with an accuracy of over 94\%. \\
The last implementation that we present in this overview comes from Bandyopadhyay et al. where they present a full Neural Network chip\cite{Bandyopadhyay2022} (fig. \ref{fig:SVD}f). The chip presents input modulators to encode the input, 3 matrix multiplication unit using the MZI mesh, interleaved by 2 nonlinear layers. The nonlinear function will be discussed in a later section. Even in this complex chip, it is possible to perform in-situ training, showing how a silicon photonic chip can cover all the tasks required by a NN. \\
The use of MZI mesh comes with several advantages, like the ideality of the MZI response (even without perfect components\cite{miller2015perfect}), the coherent scheme that requires just one single laser, and the speed of reconfigurability allowed by the pull-down p-n junction configuration of the MZI. Thanks to the reliability of the configuration and the single laser source, this approach already showed promising results and startups hit the market with solutions based on it. Moreover, even the bit resolution achieved takes advantage of this advanced state-of-the-art, reaching a high bit resolution, up to 10 bit.\\	
On the other side, this configuration comes with some limits, mainly due to the higher complexity behind SVD and the footprint required to fulfill this operation. Dividing the matrix requires a pre-computational step, as well as more components integrated into the PIC, increasing the complexity of the whole architecture. \\
In terms of component scaling and technologies, the MZM can present limitations and opportunities\cite{Alqadasi2022}. In the Reck scheme, the number of MZI needed to implement one of the two unitary matrices is $N(N-1)$, where $N$ is the number of inputs, resulting in a scaling law of $O(N^2)$. In particular, for each MZI 2 phase controls are needed (one for one input, and one for one of the arms). By pruning, the MZI required can be reduced to $N log2(N)$, under certain conditions, resulting in an important reduction of the controllers needed, as shown in fig. \ref{fig:MZM_scal}. However, to use the scheme proposed by Feng et al, the number of inputs should be a power of $2$, or the optical power unbalanced must be addressed with more MZIs.
\\
Some of the downsides of the approach using MZI mesh can be identified in the single MZI element. For example, MZI requires precise control of the phase of each path, making the phase actuator a key element in the performances, as well as being sensitive to the fabrication variability on each waveguide. Several groups analyzed the actual errors and noise due to the phase mismatch to better calibrate the impact on the NN. On the other side, some groups implemented on-chip training, forcing the same NN to calibrate itself on these errors\cite{cem2022data,9252466,9489417}.\\
Other limitations that come from the use of the MZI are the lack of parallelism and P-RAM elements. First, by using MZI, the calibration is wavelength dependent, making more challenging the implementation of a WDM-based scheme on the same MZI mesh. This lack of parallelism could limit the possibility of the architecture, relying just on the speed of the input modulators and output photodetectors. The second element is the complex integration of P-RAM components in the mesh. Those components are one of the keys for an energy-efficient PNN chip, as the PCM material they are based on, can store the weight values in a non-volatile fashion, reducing further the operation-over-energy figure of merit. However, most of the PCM materials have an impact on both amplitude and phase, making the control of one MZI more challenging. Moreover, due to the bi-level nature of the PCM, multiple strips might be required to match the offset due to fabrication phase mismatch.\\
\subsection{$ Y = M X$}
\begin{figure*}[t!]
    \centering
    \includegraphics[width=\textwidth]{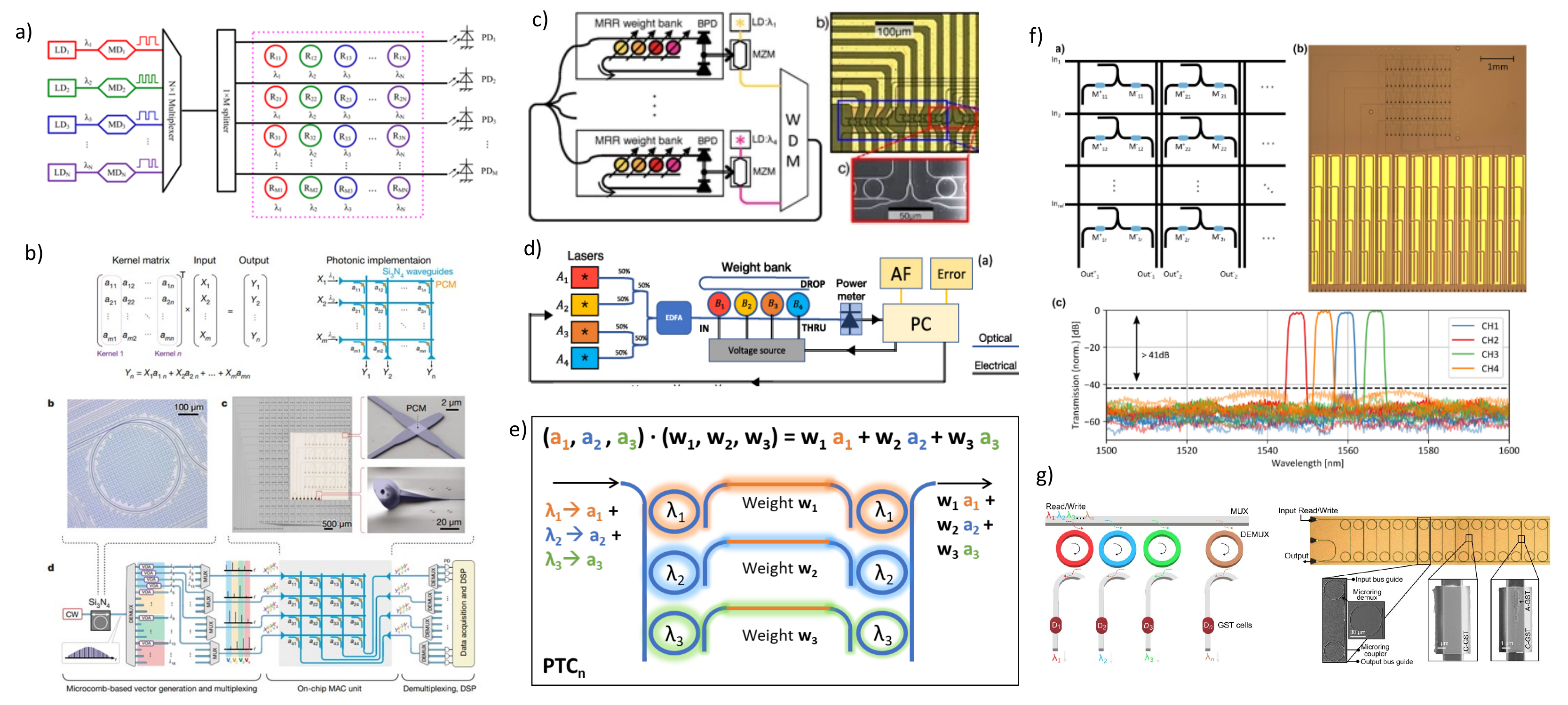}
    \caption{Architecture and PICs using WDM to implement the Matrix-Vector Multiplication. (a) First architecture proposed by Yang et al.\cite{Ji2012}, where the one-to-one matrix mapping is clearly visible. (b) Architecture exploiting cross-bar attenuated couplers, presented by Feldmann et al.\cite{Feldmann2021}. (c) First implementation of "broadcast and weight" approach from Tait et al.\cite{Tait2014}. (d) Similar "broadcast and weight" approach, that can perform training and testing of a Hopfield network\cite{Liu2021}. (e) Implementation of WDM matrix multiplication using add-drop microring resonators, implemented by Ma et al. \cite{ma2022high}. (f) Recent implementation of the cross-bar approach, improved by Bragg gratings to reduce the cross-talk\cite{Bruckerhoff2022}. (g) Add-drop PCM microring approach to demonstrate an integrated engine for unsupervised correlation detection\cite{Sarwat2022}.  }
    \label{fig:WDM}
\end{figure*}
Another approach to performing the matrix multiplication is the direct mapping of the $M$ matrix into the PIC, by exploiting one of the degrees of freedom that photonics has, such as wavelength, modes, or polarization. The most common is Wavelength-division multiplexing (WDM), where different scaled wavelength sources are combined to obtain an equivalent dot product at the photodetectors. \\
Initial architectures come from the optical computing field, where several researchers were emulating the digital logic functions of electronics\cite{Xu_11,Ibrahim2004}. The first implementation in a full WDM scheme was presented by Yang et al. in 2012\cite{Ji2012} (fig. \ref{fig:WDM}a). In this work, the matrix values are mapped one-by-one on the microring resonators grid, exploiting the MUX/DEMUX scheme for WDM lasers, where the input vector is encoded into the amplitude of the same lasers. The photodetectors at each output provide the summation of the different wavelength signals. As most of the schemes in this section, mapping the matrix $M$, the complexity of the circuit scales with the size of the matrix itself, so $O(N^2)$ for a square matrix of size $N$, but it can support rectangular matrices, as well as branch pruning to reduce the scaling factor. \\
A step forward was made by Tait et al. in 2014, describing the "broadcast and weight" approach for the optical neural network\cite{Tait2014}, later implemented in 2017\cite{Tait2017} (fig. \ref{fig:WDM}c). The architecture shows the broadcast of all the input to all the microring resonator weight banks, whose outputs are fed into the input by an amplitude modulator. The weighting is performed by tuning the microring resonators to the input wavelengths, archiving both positive and negative values thanks to the balanced photodetectors. This approach permits obtaining an optical neural network that has been demonstrated useful for many applications\cite{huang2021silicon}. Other implementations have exploited the tunability of add-drop microring resonators as weights to perform the multiplication as attenuation of the incoming light beam\cite{Liu2021} (fig. \ref{fig:WDM}d), reaching up to 9 bit resolution\cite{Zhang2022}. The use of microrings allows for an important footprint reduction (using SiPh, microring could be downsized to a 10 $\mu$m radius) while having high-speed reconfiguration thanks to the internal p-i-n junction, that nowadays could reach a bandwidth of more than 25 GHz. Moreover, thanks to the add-drop configuration, the architecture could have both positive and negative sign weights in the matrix, without the need for post-processing to correct the data. The main disadvantage is coming from the control perspective, as microring tends to be a sensitive element towards noise sources, such as temperature variation, stress, and so on. By so, besides the modulation controlling the p-i-n junction, another signal must be applied to the heater to assure a perfect alignment between the microring’s resonance and the laser’s wavelength, doubling the number of controls. Moreover, due to this high integration and need for resonance stabilization, integration of P-RAM elements in the ring itself is challenging due to the double $n-k$ impact of the material and the finite number of states, making this architecture not directly suitable for low-energy applications, such as edge computing.\\
Another approach exploits tunable couplers between rows and columns of an optical waveguide grid, presented by Feldmann et al. in 2021\cite{Feldmann2021} (fig. \ref{fig:WDM}b). Each wavelength coming from a Comb laser source is modulated and fed into a certain row. The tunable couplers bend a certain amount of the incoming light toward the selected column. The photodiode at the end of the column collects the composition of the different light beams, whose amplitude is determined by the couplers and the P-RAM element placed after the coupler. This scheme relies on the simplicity of the implementation that reduces the number of controllers to the minimum (equal to the size of the matrix), and implementing them with PCM allows having almost 0 energy cost, but limits the speed of reconfiguration. A further improvement was presented in 2022\cite{Bruckerhoff2022}, where Bragg gratings are used to reduce the crosstalk between channels, and so improving the resolution (fig. \ref{fig:WDM}f). \\
The last architecture was presented by Miscuglio et al.\cite{Miscuglio2020}, and later implemented by Ma et al.\cite{ma2022high} (fig. \ref{fig:WDM}e). This architecture takes advantage of the add-drop microring as the element to fan-out the WDM inputs and recombines them after attenuation is applied in the waveguide link between them. This approach has the advantage to be able to use PCM, slow-speed heater-based components, and high-speed p-i-n junction to achieve the required attenuation, by so fulfilling the requirement of both edge computing applications and cloud one. The number of controls could be high in principle (up to 3 controllers for each element of the matrix), however, by relying on the fabrication quality and accepting a reduction of the resolution, the control could be reduced to just an attenuator per element of the matrix. Sarwat et al. used the same approach to demonstrate an integrated engine for unsupervised correlation detection\cite{Sarwat2022} (fig. \ref{fig:WDM}g).\\
Similar architectures can be implemented by exploiting mode or polarization multiplexing or mixing different approaches for more compact and yet performance implementations. The mapping of the weight matrix in a one-to-one fashion allows to have a higher level of flexibility, and requires less pre-computation, as it does not require any matrix decomposition. However, the scaling factor will follow the size of the matrix itself, posing an important limitation due to the high number of components required, and the control electronics circuits they require. \\ 
\section{Architectures Comparison}
\begin{table*}[t!]
\caption{Scaling comparison of various approaches to performing MVM and MAC operations using photonic chip-based components. N = size of input vector; M = size output vector; P-RAM = Photonic random access memory, allowing for zero-static power consumption, once the weights are SET.}
\label{table_1}
\centering
\resizebox{\textwidth}{!}{
\begin{tabular}{|c||c||c||c||c|}
\hline
 \multirow{2}{*}{Type of Operation} & $Y = V^T \Sigma U X$ & \multicolumn{3}{|c|}{ $ Y = M X$} \\
 & \includegraphics[height=0.5cm]{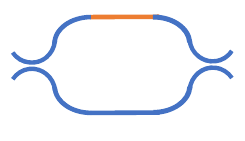} & \includegraphics[height=0.5cm]{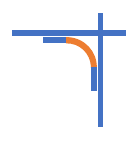} & \includegraphics[height=0.5cm]{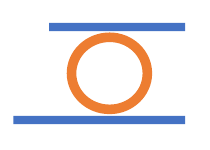} & \includegraphics[height=0.5cm]{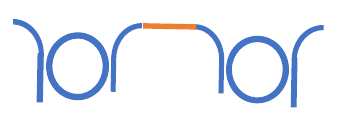}\\
\hline
\# Input & 1 Laser, N Modulators & 1 Comb Laser, N Modulators & N Lasers, N Modulators & N Lasers, N Modulators\\
\hline
\# Outputs & N Photodiodes & M Photodiodes & M or $2*$M Photodiodes & M Photodiodes\\
\hline
Area/Basic Element Area & $N^2-N$ or $N log_2(N)$ & $N$×$M$ & $N$×$M$ & $N$×$M$ \\
\hline
Controllers & $2N^2-N$ or $N log_2(N)$ & $N$×$M$ & $2(N$×$M)$ & $N$×$M$\\
\hline
Parallelization & No & WDM Off Chip & WDM On Chip & WDM On Chip\\
\hline
Weight Bit Resolution & 8/10 & 5 & 9 & $>$5\\
\hline
P-RAM & No & Yes & No & Yes \\
\hline
\end{tabular}}
\end{table*}
As seen, many different architectures could be used to implement MVM for neural networks, as summarized in table \ref{table_1}. In the table, actual Figure-of-Merit MAC/s nor MAC/J is not reported, as for all the architectures, it will mainly rely on the inputs modulator and output photodiodes, whose characteristics are coming from the fabrication process rather than the component used to perform the MAC operation. However, in case where weights must be updated at the same speed of the inputs, the architecture choice will reduce to the ones that allow an high-speed weight updates (for example using p-n junctions), to respect to slow or large footprint ones. \\
One parameter that influences the choice of architecture is the chip footprint, based on the size of the component used and the scaling compared to the matrix. The basic  $Y= M X$ architecture  takes advantage of the more direct equation, as scaling is proportional to the matrix size, while the MZM approach suffers from the decomposition matrices. However, for both approaches, the scaling follows $O(N^2)$, except for the butterfly configuration used by Feng et al. This last one exploits pruning as a way to reduce the number of connections, and so the scaling of the circuit. For the number of controls, the best solutions appear to be the one based on couplers and coupled microrings, even if this last one might be affected by the detuning of the microrings that would limit the Extinction Ratio, and so the bit resolution achievable by the NN. The architecture based on single add-drop microrings could either have the same $M$×$N$ controllers if just one tuning method is used (for example employing just heaters as tuning weight), but each microring needs to integrate both a trimming method (i.e. heaters) and a high-speed tuning (i.e. the p-i-n junction) to support high-speed reconfiguration, doubling the number of controls. The lack of need for tuning for the coupler architecture comes at the cost of a more complex input that requires a comb laser, and a WDM mux and demux external to the chip for the output, increasing the complexity of the overall system. \\
Bit resolution shows a strong point for the architectures based on MZM, for mainly two reasons: the more straightforward capability of controlling the phase difference in the MZM, resulting in a larger ER, and so larger bit resolution; the advanced stage of the products based on this technology that already reached the market, so having passed the optimization process. Different types of modulations, for example based on Electro-Adsorption Modulators\cite{amin2018waveguide}, can provide a higher ER in a compact way, allowing a high bit resolution also for other architecture. Moreover, techniques such as coherent detection have been proposed\cite{Zhang2022}, capable to reach 9-bit resolution with WDM MRR architecture.   \\
The last piece of confrontation is regarding the possibility to implement P-RAM on the circuits\cite{rios2015integrated, cheng2018device, alexoudi2020optical, Peserico2022}, by using PCMs for example\cite{meng2022electrical}. In a larger view, as more and more MVM circuits will be used to implement NNs, having the possibility to integrate photonic memory elements would have a crucial benefit in terms of energy efficiency, as it reduces the power needed to tune the weight as well as the energy required to access external memory elements in DRAM\cite{petrenko2018limitations}. That would allow targeting edge computing applications, rather than just cloud applications in data centers, where power consumption is a priority to extend the lifespan of those devices. Up to our knowledge, just two architectures allow the integration of the PCMs, placing those materials either in the couplers or between coupled rings. The architecture based on MZM could benefit in case a phase-only PCM would be presented, as most of the materials are now affecting adsorption too, such as GSST\cite{sahoo2022gsst}, GSSe\cite{meng2022electrical}, or GST\cite{redaelli2022material}. Integration of PCMs into microring resonator might be challenging for the same reason, adding also a problem of cross-heating interference, as the tuning element could affect the phase of the material, resulting in an unwanted switching.

\section{Nonlinearities}
\begin{figure*}[t!]
    \centering
    \includegraphics[width=\textwidth]{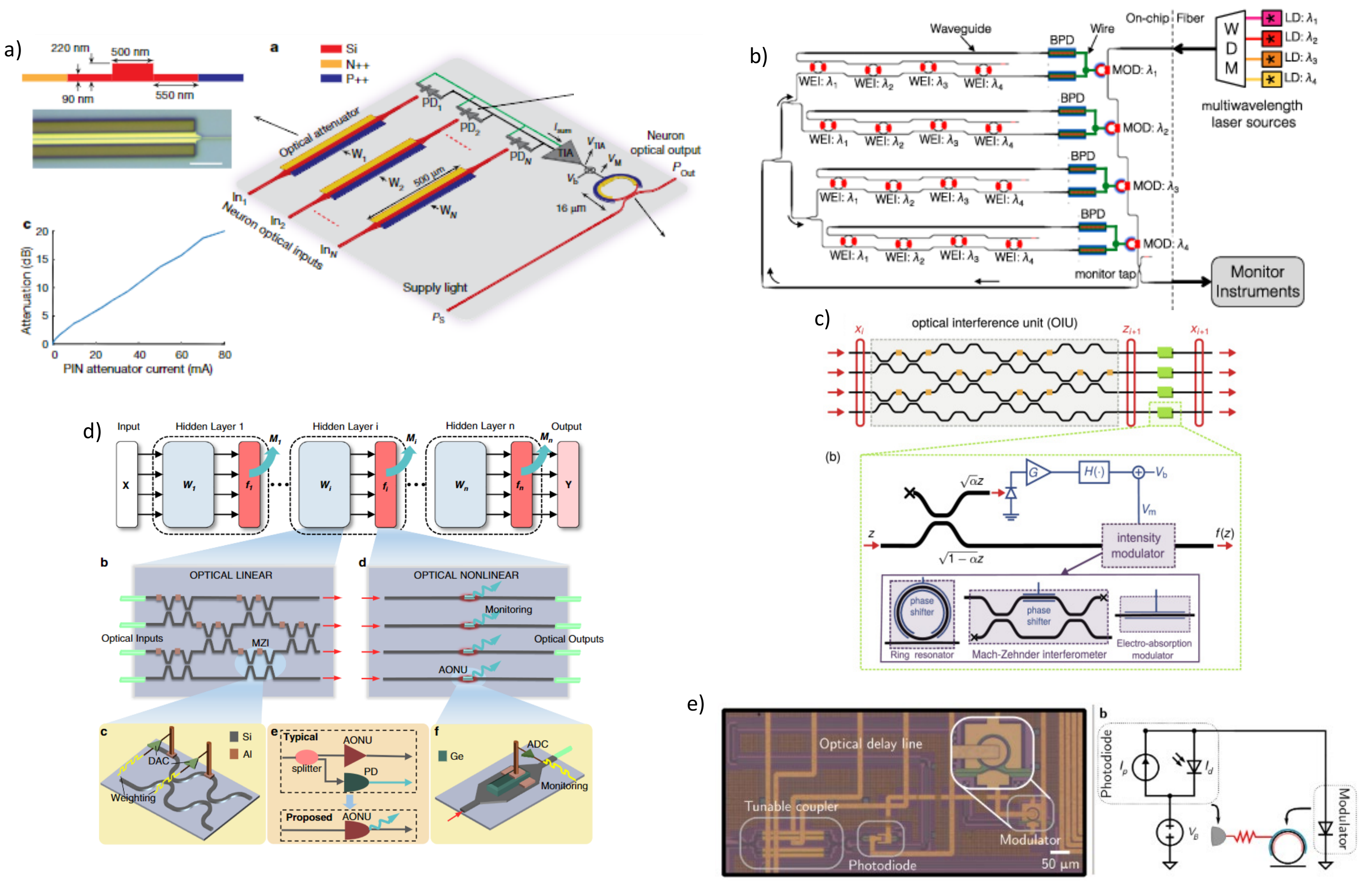}
    \caption{Circuits and architectures implementing the activation function on-chip. (a) Ashtian et al. show a full O/E/O domain change to pilot a MRR as nonlinear response\cite{Ashtiani2022}. (b) Tait et al. use a similar approach to pilot an add-drop microrings bank, using balanced photodetectors\cite{tait2019silicon}. (c) Experimental implementation of arbitrary activation function by tapping part of the optical output power\cite{Moayedi2020}. (d) Shi et al. present the use of a short SiGe photodetetors to implement a nonlinear transfer function between optical power input and output\cite{Shi2022}. (e) Implementation used by Bandyopadhyay et al., where just a tap of the optical output is used to modulate itself in nonlinear way by a MRR\cite{Bandyopadhyay2022}. } 
    \label{fig:NL}
\end{figure*}
The last piece to turn a PIC circuit performing MVM into a NN is by providing a nonlinear activation function. In many of the NNs we have seen before, this activation function was performed by a CPU or GPU, once the optical signal is transformed into a digital one. This conversion allows several advantages, like performing mathematical complex functions (including calibration), as well as having the flexibility to change the actual activation function based on the goals of the NN. However, it presents several drawbacks: one is the slow speed associated with this procedure, linked with the long latency, that nullify the major advantages of implementing a photonic neural networks. The power involved is also a major drawback: it has been demonstrated that ADCs are the first contributes in the energy cost of the system, especially for high speed ones\cite{gupta20224f}. Moreover, to perform the following neural layer, all the digitized signals must converted back into optical analog domain, requiring additional energy. \\
To overcome these limitations, and so keep the high bandwidth and low latency provided by the optical domain, many researchers have explored different solutions. One of the major paths is the exploiting of material nonlinearities on-chip, which can be exploited by high-power optical signals under certain conditions. While this path comes from a long tradition of exploring nonlinearities in silicon or silicon nitride waveguide (for generating single photons\cite{leuthold2010nonlinear}, four-wave mixing\cite{morichetti2011travelling}, or comb generation\cite{chang2022integrated}), the cost of dealing with high power signals limit the possibility of implementation into large and deep optical neural network at the moment.\\
By so, other architectures have been explored to implement such solutions, that use, completely or partially, an electrical-optical domain change, while keeping the signal in the analog domain. Here, we list the major ones, based on the approach used.
\subsection{Full O/E/O Conversion}
The first implementation we present is the complete conversion of the optical signal into an electrical one, that would latter pilots a novel optical signal. One implementation presented by Ashtian et al.\cite{Ashtiani2022} (fig. \ref{fig:NL}a), and similarly by Tait et al.\cite{tait2019silicon} (fig. \ref{fig:NL}b), implements the activation function by modulating  the wavelength resonance of a microring resonator, that is fed from an external CW laser source that can be directly sent into the following neural layer. The 2 implementations have some differences: Ashtian et al. implement the summation by combining the current of the photodetectors, directly connected to the modulator and inputs. Moreover, a stage of amplification is placed to match the voltage levels between the sum of the photocurrents and the p-n junction of the microring resonator. Using this scheme, there is no need for WDM multiplexing. On the other hand, Tait et al. use a WDM scheme in loop-back with differential photodetectors to tune directly the microring resonators. This scheme presents some limitations and opportunities, in particular, can be sensitive to fabrication differences between photodetectors pairs, unbalancing the actual response of the microring. This fully O/E/O approach, where a complete domain change, from optical to electrical and back to optical is used, takes advantage of the full bandwidth of the components as the signals stay in the analog domain. However, the O/E/O approach adds noise sources, in particular due to the photodetectors and amplification stage\cite{Thomas2020Noise}. To reduce the actual noise, one proposed solution is using modulators that require a lower $V_\pi L$, to reduce or completely avoid the amplification stage. Heterogeneously integrated devices, such as ITO-based modulators\cite{amin2019ito, gui2022100}, or  ITO-graphene device\cite{amin2021ito}, can reduce the $V_\pi L$ by orders of magnitude, reducing drastically the need for amplification stages, and so the noise introduced by them.\\
\subsection{Nonlinear Adsorption Devices}
Another architecture to implement a nonlinearity is by design a custom nonlinear device, so providing a nonlinear function between the optical input and output. A solution based on SiGe photodetector has been proposed by Shi et al.\cite{Shi2022}, leveraging the short structure of the SiGe to limit the maximum optical power output from the component, by so implementing a nonlinear transfer function (fig. \ref{fig:NL}d). This solution has the advantages to permit the monitoring of the power while preserving the latency of the optical circuit. Similar approaches have been explored by other groups, to find the best material to perform this function both on the detection and modulation side, aiming for a better energy-efficient way\cite{amin2019ito,feldmann2019all}. This type of approach has the potential to leverage on the different types of material that can be used, limited by the compatibility with the SiPh CMOS process. The main drawback can be identified in the scaling limitations, since the input power must meet a certain critaria to activate the nonlinear function, a large NN or a high loss PIC would not be suitable for this approach, unless other adjustments (like on-chip amplification stage) would be addopted. \\
\subsection{Light Splitting and Detection}
Another approach has been used by Moayedi et al.\cite{Moayedi2020}, and Bandyopadhyay et al.\cite{Bandyopadhyay2022}, shown respectively in fig. \ref{fig:NL}c-e. In this case, the linear part of the NN is based on the MZI mesh, and the nonlinear function is activated by just part of the light power of the output waveguide (splitting), which is detected and the signal used to modulate the amplitude of the remaining part of the optical signal.
This allows feeding the whole network with the same optical input signals, reducing the need to have more lasers or input couplers. However, the network will add layers of modulation on top of each other, making the same scheme more sensitive to noise and not directly suitable for WDM expansions. Following the same approach, Xu et al.\cite{Xu2022} propose a similar scheme. In this case the NL part is implemented using a MZI, where one of the arm is controlled by a optical memory-based feedback circuit, using a PCM material as nonvolatile element. The light-splitting-and-detection has some clear advantages as the modulation is directly on the same optical signal, with a clear advantage in terms of speed and latency. However, the tap requires an electrical circuit capable of reading low currents and translate into proper signals, limiting the actual bandwidth, and posing limitation in the energy consumption as well. Moreover, the continuous splitting layer after layer increases the insertion loss of the overall photonic circuit.\\ \\
In all cases, the activation function is encoded at the hardware level, resulting in a fixed size of the number of inputs, layers, and outputs. Schemes that can be used to subdivide the matrix into smaller ones to fit large MVM on smaller hardware cannot be used in this scheme, as the nonlinear function is applied \emph{a-priori}. By so, the research may look into schemes that allow an actual flexible implementation of the nonlinear function, by exploiting more programmable photonic circuits for example. \\

\section{Discussion}
In all this review, it has been clear that photonics has a great opportunity to be the hardware accelerator for NN applications, as the increasing number of machine learning applications is driving the actual hardware to its limit. Integrated photonics, and Silicon Photonics (SiPh) as the main actor, have several advantages and directions that could drive the implementation of fast and reliable Photonic Neural Networks (PNNs). The research and progress that have been done in the last decades for mainly telecom purposes have now a new shine in another field. Among them, we can see the main driving forces: 
\begin{itemize}
    \item \emph{Components}: SiPh can now show several components that are over the possibilities of any electrical counterpart in terms of speed and energy efficiency. Modulators up to 100 GHz\cite{rahim2021taking}, and photodetectors that can reach over 200 GHz have been presented\cite{lischke2021ultra}, while CMOS foundries are more and more implementing SiPh lines, with state-of-the-art components in their PDKs. Note that all these components come from another field (telecom mainly), but their impact can be further beyond the initial field. 
    \item \emph{Emerging Materials and Devices}: The research over new materials and new devices has brought several innovations in recent years\cite{Peserico2022, fratalocchi2015nano}. Among many, ITO has shown the most potential, especially in terms of energy efficiency and footprint, being 1000x smaller than Si EOM, and 10,000x smaller than Lithium Niobate\cite{amin2019ito, amin2021ito, gui2022100, wang2022high, amin2020sub, pickus2013silicon}. Beyond ITO, two-dimensional material-based solutions may yet play a role in future semiconductor chips and tensor core processors; for instance, the accumulation operation can be performed simply and incoherently using a photodetector as discussed above. The detector's figure of merit, the gain-bandwidth-product, falls into either sensitive-but-slow or into fast-but-non-sensitive quadrants \cite{sorger2020roadmap}. Recent developments on slot-based 2D material detectors show to overcome the transient-time and RC-delay time limitations offering sensitive and fast detectors while offering a minuscule footprint. Furthermore, PN-junction-based 2D detectors have demonstrated promising performance while not requiring a bias, thus saving power and wire-routing complexity \cite{patil2022highly, wang2022self}.
    \item \emph{I/O}: One of the limitations that slowed down the expansion of SiPh was the actual coupling in/out of the chip, an essential piece considering also the lack of integrated light sources. SiPh has now advanced packaging tools to provide small form factor chips, with laser sources on-chip\cite{wang2017novel}. Moreover, the expansion of the materials used has brought new devices, such as P-RAM\cite{meng2022electrical}, to be integrated, reducing the dependency on external digital electronic memories, one of the bottlenecks of electronics. The next steps will focus on inter-chip communications, as well as intra-chip ones. 
    \item \emph{Domain Crossings}: Photonic-based tensor core processors are analog in nature and hence may require digital to analog and vise versa domain crossings. Above 5GHz baud rates and 8-bit resolution DACs and ADCs become quite expensive to operate. If the PTC application allows processing data in the optical domain (from an optical input, such as for intra data-center, for example), then a photonic PIC-based DAC would be beneficial \cite{meng2021electronic}. Energy harvesting such as recapturing optical nonlinearities \cite{jalali2006raman} or nanoscale RF antennas or solar cells \cite{tahersima2015enhanced}.
    \item \emph{Architectures}: As seen, several architectures have been proposed and demonstrated. While a clear winner is still to be found, all of them can push towards several improvements to further expand their performances\cite{sunny2021survey}. On one side, parallelization exploiting other degrees of freedom can further push the performances. On the other side, techniques such as pruning or others can be implemented on-chip as well, making room for improvements in the overall system. There may also be options to learn from emerging architectures such as hybrid (electronic photonic) network-on-chip approaches that allocate interconnect technology between local (electronic) and distant (optical) requirements, which may also allow for some degree of network reconfiguration for demand optimization \cite{narayana2017morphonoc, shen2022reconfigurable}.
\end{itemize}
However, photonics has still to improve some aspects to become a viable solution for deep learning and machine intelligence. 
Adding a nonlinear activation function in the optical domain is challenging, and more efficient all-optical nonlinearities need to be explored, yet, electro-optical nonlinearity devices are promising, despite some architectural overhead, such footprint, accumulation detectors, and ADCs. 
Analog-to-digital and digital-to-analog conversions must be taken into account too: domain crossings (i.e. DAC \& ADCs) constitute the majority power consumption for heterogeneous photonic-electronic machine intelligence accelerators. However, emerging monolithic integration  solutions (e.g. Global Foundry 45nm, GF45SPLCO\cite{9083571}) hold great promise to minimized communication overhead. Furthermore, emerging packaging solutions including stacking multiple BOELs\cite{chan2016beol, standaert2016beol}, integrating plurality of chiplets onto a same interposer, with world-record pin pitches  of 10$\mu$m\cite{vega2019three}, will enable extremely tight integrated heterogeneous PNN-CMOS ASIC solution with unprecedented performance. The upcoming SRC JUMP2.0 Center will  explore the latter in detail. 
By last, laser integration must become a standard in the SiPh process, allowing high energy-efficient lasers to be implemented on-chip, exploring integration\cite{wang2017novel}, or Photonic Wire Bonding\cite{billah2018hybrid}. \\
\section{Conclusions}
In this paper, we review the main aspects that enable integrated photonic technologies to become a key resource for the current and next generation of hardware accelerators for Neural Networks. We review the main advantages that photonics has compared to electronics, in terms of power efficiency, latency, and bandwidth. The main architectures that have been used so far to implement linear and nonlinear operations on PIC have been shown, highlighting the pros and cons of each one of them, and outlining a comparison among them. We finally discuss among the main drive forces that will boost the photonic approach in the next years. \\
Considering all those aspects, photonics will still play an important role in the research for the next generation of hardware accelerators. As more and more computational power is required and considering energy efficiency a key factor, photonics will be in the spotlight in the near and long-term future.

\appendices


\section*{Acknowledgment}

V.J.S. is supported by the PECASE Award under the AFOSR grant (FAA9550-20-1-0193)

\ifCLASSOPTIONcaptionsoff
  \newpage
\fi



%
\bibliographystyle{IEEEtran}
\bibliography{refs}

\begin{thebibliography}{100}
\providecommand{\url}[1]{#1}
\csname url@samestyle\endcsname
\providecommand{\newblock}{\relax}
\providecommand{\bibinfo}[2]{#2}
\providecommand{\BIBentrySTDinterwordspacing}{\spaceskip=0pt\relax}
\providecommand{\BIBentryALTinterwordstretchfactor}{4}
\providecommand{\BIBentryALTinterwordspacing}{\spaceskip=\fontdimen2\font plus
\BIBentryALTinterwordstretchfactor\fontdimen3\font minus
  \fontdimen4\font\relax}
\providecommand{\BIBforeignlanguage}[2]{{%
\expandafter\ifx\csname l@#1\endcsname\relax
\typeout{** WARNING: IEEEtran.bst: No hyphenation pattern has been}%
\typeout{** loaded for the language `#1'. Using the pattern for}%
\typeout{** the default language instead.}%
\else
\language=\csname l@#1\endcsname
\fi
#2}}
\providecommand{\BIBdecl}{\relax}
\BIBdecl

\bibitem{sarker2021machine}
I.~H. Sarker, ``Machine learning: Algorithms, real-world applications and
  research directions,'' \emph{SN Computer Science}, vol.~2, no.~3, pp. 1--21,
  2021.

\bibitem{5392560}
A.~L. Samuel, ``Some studies in machine learning using the game of checkers,''
  \emph{IBM Journal of Research and Development}, vol.~3, no.~3, pp. 210--229,
  1959.

\bibitem{fradkov2020early}
A.~L. Fradkov, ``Early history of machine learning,'' \emph{IFAC-PapersOnLine},
  vol.~53, no.~2, pp. 1385--1390, 2020.

\bibitem{yadav2015history}
N.~Yadav, A.~Yadav, and M.~Kumar, ``History of neural networks,'' in \emph{An
  introduction to neural network methods for differential equations}.\hskip 1em
  plus 0.5em minus 0.4em\relax Springer, 2015, pp. 13--15.

\bibitem{Leonard1990}
J.~Leonard and M.~Kramer, ``Improvement of the backpropagation algorithm for
  training neural networks,'' \emph{Computers \& Chemical Engineering},
  vol.~14, pp. 337--341, 3 1990.

\bibitem{mahesh2020machine}
B.~Mahesh, ``Machine learning algorithms-a review,'' \emph{International
  Journal of Science and Research (IJSR).[Internet]}, vol.~9, pp. 381--386,
  2020.

\bibitem{schmidhuber2015deep}
J.~Schmidhuber, ``Deep learning in neural networks: An overview,'' \emph{Neural
  networks}, vol.~61, pp. 85--117, 2015.

\bibitem{abiodun2018state}
O.~I. Abiodun, A.~Jantan, A.~E. Omolara, K.~V. Dada, N.~A. Mohamed, and
  H.~Arshad, ``State-of-the-art in artificial neural network applications: A
  survey,'' \emph{Heliyon}, vol.~4, no.~11, p. e00938, 2018.

\bibitem{rawat2017deep}
W.~Rawat and Z.~Wang, ``Deep convolutional neural networks for image
  classification: A comprehensive review,'' \emph{Neural computation}, vol.~29,
  no.~9, pp. 2352--2449, 2017.

\bibitem{ramesh2022hierarchical}
A.~Ramesh, P.~Dhariwal, A.~Nichol, C.~Chu, and M.~Chen, ``Hierarchical
  text-conditional image generation with clip latents,'' \emph{arXiv preprint
  arXiv:2204.06125}, 2022.

\bibitem{hwang2018computational}
T.~Hwang, ``Computational power and the social impact of artificial
  intelligence,'' \emph{arXiv preprint arXiv:1803.08971}, 2018.

\bibitem{erickson2017toolkits}
B.~J. Erickson, P.~Korfiatis, Z.~Akkus, T.~Kline, and K.~Philbrick, ``Toolkits
  and libraries for deep learning,'' \emph{Journal of digital imaging},
  vol.~30, no.~4, pp. 400--405, 2017.

\bibitem{reuther2019survey}
A.~Reuther, P.~Michaleas, M.~Jones, V.~Gadepally, S.~Samsi, and J.~Kepner,
  ``Survey and benchmarking of machine learning accelerators,'' in \emph{2019
  IEEE high performance extreme computing conference (HPEC)}.\hskip 1em plus
  0.5em minus 0.4em\relax IEEE, 2019, pp. 1--9.

\bibitem{suda2016throughput}
N.~Suda, V.~Chandra, G.~Dasika, A.~Mohanty, Y.~Ma, S.~Vrudhula, J.-s. Seo, and
  Y.~Cao, ``Throughput-optimized opencl-based fpga accelerator for large-scale
  convolutional neural networks,'' in \emph{Proceedings of the 2016 ACM/SIGDA
  International Symposium on Field-Programmable Gate Arrays}, 2016, pp. 16--25.

\bibitem{zhou2017tunao}
J.~Zhou, S.~Liu, Q.~Guo, X.~Zhou, T.~Zhi, D.~Liu, C.~Wang, X.~Zhou, Y.~Chen,
  and T.~Chen, ``Tunao: A high-performance and energy-efficient reconfigurable
  accelerator for graph processing,'' in \emph{2017 17th IEEE/ACM International
  Symposium on Cluster, Cloud and Grid Computing (CCGRID)}.\hskip 1em plus
  0.5em minus 0.4em\relax IEEE, 2017, pp. 731--734.

\bibitem{reuther2020survey}
A.~Reuther, P.~Michaleas, M.~Jones, V.~Gadepally, S.~Samsi, and J.~Kepner,
  ``Survey of machine learning accelerators,'' in \emph{2020 IEEE high
  performance extreme computing conference (HPEC)}.\hskip 1em plus 0.5em minus
  0.4em\relax IEEE, 2020, pp. 1--12.

\bibitem{han2015learning}
S.~Han, J.~Pool, J.~Tran, and W.~Dally, ``Learning both weights and connections
  for efficient neural network,'' \emph{Advances in neural information
  processing systems}, vol.~28, 2015.

\bibitem{Shastri2021}
B.~J. Shastri, A.~N. Tait, T.~F. de~Lima, W.~H. Pernice, H.~Bhaskaran, C.~D.
  Wright, and P.~R. Prucnal, ``Photonics for artificial intelligence and
  neuromorphic computing,'' \emph{Nature Photonics}, vol.~15, pp. 102--114, 2
  2021.

\bibitem{ma2021deep}
W.~Ma, Z.~Liu, Z.~A. Kudyshev, A.~Boltasseva, W.~Cai, and Y.~Liu, ``Deep
  learning for the design of photonic structures,'' \emph{Nature Photonics},
  vol.~15, no.~2, pp. 77--90, 2021.

\bibitem{Miscuglio2020}
\BIBentryALTinterwordspacing
M.~Miscuglio and V.~J. Sorger, ``Photonic tensor cores for machine learning,''
  \emph{Applied Physics Reviews}, vol.~7, p. 031404, 2 2020. [Online].
  Available: \url{http://arxiv.org/abs/2002.03780
  http://dx.doi.org/10.1063/5.0001942}
\BIBentrySTDinterwordspacing

\bibitem{miller2017attojoule}
D.~A. Miller, ``Attojoule optoelectronics for low-energy information processing
  and communications,'' \emph{Journal of Lightwave Technology}, vol.~35, no.~3,
  pp. 346--396, 2017.

\bibitem{sunny2021survey}
F.~P. Sunny, E.~Taheri, M.~Nikdast, and S.~Pasricha, ``A survey on silicon
  photonics for deep learning,'' \emph{ACM Journal of Emerging Technologies in
  Computing System}, vol.~17, no.~4, pp. 1--57, 2021.

\bibitem{soref1987electrooptical}
R.~Soref and B.~Bennett, ``Electrooptical effects in silicon,'' \emph{IEEE
  journal of quantum electronics}, vol.~23, no.~1, pp. 123--129, 1987.

\bibitem{chrostowski2015silicon}
L.~Chrostowski and M.~Hochberg, \emph{Silicon photonics design: from devices to
  systems}.\hskip 1em plus 0.5em minus 0.4em\relax Cambridge University Press,
  2015.

\bibitem{siew2021review}
S.~Y. Siew, B.~Li, F.~Gao, H.~Y. Zheng, W.~Zhang, P.~Guo, S.~W. Xie, A.~Song,
  B.~Dong, L.~W. Luo \emph{et~al.}, ``Review of silicon photonics technology
  and platform development,'' \emph{Journal of Lightwave Technology}, vol.~39,
  no.~13, pp. 4374--4389, 2021.

\bibitem{haensch2017scaling}
W.~Haensch, ``Scaling is over—what now?'' in \emph{2017 75th Annual Device
  Research Conference (DRC)}.\hskip 1em plus 0.5em minus 0.4em\relax IEEE,
  2017, pp. 1--2.

\bibitem{etiemble201845}
D.~Etiemble, ``45-year cpu evolution: one law and two equations,'' \emph{arXiv
  preprint arXiv:1803.00254}, 2018.

\bibitem{petrenko2018limitations}
S.~Petrenko, ``Limitations of von neumann architecture,'' in \emph{Big Data
  Technologies for Monitoring of Computer Security: A Case Study of the Russian
  Federation}.\hskip 1em plus 0.5em minus 0.4em\relax Springer, 2018, pp.
  115--173.

\bibitem{ganguly2019towards}
A.~Ganguly, R.~Muralidhar, and V.~Singh, ``Towards energy efficient non-von
  neumann architectures for deep learning,'' in \emph{20th international
  symposium on quality electronic design (ISQED)}.\hskip 1em plus 0.5em minus
  0.4em\relax IEEE, 2019, pp. 335--342.

\bibitem{cheng2017survey}
Y.~Cheng, D.~Wang, P.~Zhou, and T.~Zhang, ``A survey of model compression and
  acceleration for deep neural networks,'' \emph{arXiv preprint
  arXiv:1710.09282}, 2017.

\bibitem{deng2020model}
L.~Deng, G.~Li, S.~Han, L.~Shi, and Y.~Xie, ``Model compression and hardware
  acceleration for neural networks: A comprehensive survey,'' \emph{Proceedings
  of the IEEE}, vol. 108, no.~4, pp. 485--532, 2020.

\bibitem{Ding2017}
\BIBentryALTinterwordspacing
C.~Ding, S.~Liao, Y.~Wang, Z.~Li, N.~Liu, Y.~Zhuo, C.~Wang, X.~Qian, Y.~Bai,
  G.~Yuan, X.~Ma, Y.~Zhang, J.~Tang, Q.~Qiu, X.~Lin, and B.~Yuan, ``Circnn:
  Accelerating and compressing deep neural networks using block-circulant
  weight matrices; circnn: Accelerating and compressing deep neural networks
  using block-circulant weight matrices,'' \emph{2017 50th Annual IEEE/ACM
  International Symposium on Microarchitecture (MICRO)}, vol.~14, 2017,
  13-17<br/>21,22<br/>23, 28-33<br/><br/>DRAM 34,35. [Online]. Available:
  \url{https://doi.org/10.1145/3123939.3124552}
\BIBentrySTDinterwordspacing

\bibitem{han2015deep}
S.~Han, H.~Mao, and W.~J. Dally, ``Deep compression: Compressing deep neural
  networks with pruning, trained quantization and huffman coding,'' \emph{arXiv
  preprint arXiv:1510.00149}, 2015.

\bibitem{sung2015resiliency}
W.~Sung, S.~Shin, and K.~Hwang, ``Resiliency of deep neural networks under
  quantization,'' \emph{arXiv preprint arXiv:1511.06488}, 2015.

\bibitem{blalock2020state}
D.~Blalock, J.~J. Gonzalez~Ortiz, J.~Frankle, and J.~Guttag, ``What is the
  state of neural network pruning?'' \emph{Proceedings of machine learning and
  systems}, vol.~2, pp. 129--146, 2020.

\bibitem{astrid2018deep}
M.~Astrid and S.-I. Lee, ``Deep compression of convolutional neural networks
  with low-rank approximation,'' \emph{ETRI journal}, vol.~40, no.~4, pp.
  421--434, 2018.

\bibitem{zhuang2018towards}
B.~Zhuang, C.~Shen, M.~Tan, L.~Liu, and I.~Reid, ``Towards effective
  low-bitwidth convolutional neural networks,'' in \emph{Proceedings of the
  IEEE conference on computer vision and pattern recognition}, 2018, pp.
  7920--7928.

\bibitem{pethick2003parallelization}
M.~Pethick, M.~Liddle, P.~Werstein, and Z.~Huang, ``Parallelization of a
  backpropagation neural network on a cluster computer,'' in
  \emph{International conference on parallel and distributed computing and
  systems (PDCS 2003)}, 2003.

\bibitem{saiyeda2017cloud}
A.~Saiyeda and M.~A. Mir, ``Cloud computing for deep learning analytics: A
  survey of current trends and challenges.'' \emph{International Journal of
  Advanced Research in Computer Science}, vol.~8, no.~2, 2017.

\bibitem{yang2017method}
T.-J. Yang, Y.-H. Chen, J.~Emer, and V.~Sze, ``A method to estimate the energy
  consumption of deep neural networks,'' in \emph{2017 51st asilomar conference
  on signals, systems, and computers}.\hskip 1em plus 0.5em minus 0.4em\relax
  IEEE, 2017, pp. 1916--1920.

\bibitem{rani2021survey}
R.~Rani and R.~Garg, ``A survey of thermal management in cloud data centre:
  Techniques and open issues,'' \emph{Wireless Personal Communications}, vol.
  118, no.~1, pp. 679--713, 2021.

\bibitem{machupalli2022review}
R.~Machupalli, M.~Hossain, and M.~Mandal, ``Review of asic accelerators for
  deep neural network,'' \emph{Microprocessors and Microsystems}, vol.~89, p.
  104441, 2022.

\bibitem{zhang2018analyzing}
J.~J. Zhang, T.~Gu, K.~Basu, and S.~Garg, ``Analyzing and mitigating the impact
  of permanent faults on a systolic array based neural network accelerator,''
  in \emph{2018 IEEE 36th VLSI Test Symposium (VTS)}.\hskip 1em plus 0.5em
  minus 0.4em\relax IEEE, 2018, pp. 1--6.

\bibitem{Burgess19nvidia}
\BIBentryALTinterwordspacing
J.~Burgess, ``{RTX} {ON} - the {NVIDIA} {TURING} {GPU},'' in \emph{2019 {IEEE}
  Hot Chips 31 Symposium (HCS), Cupertino, CA, USA, August 18-20, 2019}.\hskip
  1em plus 0.5em minus 0.4em\relax {IEEE}, 2019, pp. 1--27. [Online].
  Available: \url{https://doi.org/10.1109/HOTCHIPS.2019.8875651}
\BIBentrySTDinterwordspacing

\bibitem{Yang19Intel}
\BIBentryALTinterwordspacing
A.~Yang, ``Deep learning training at scale spring crest deep learning
  accelerator (intel{\textregistered} nervana{\texttrademark} {NNP-T)},'' in
  \emph{2019 {IEEE} Hot Chips 31 Symposium (HCS), Cupertino, CA, USA, August
  18-20, 2019}.\hskip 1em plus 0.5em minus 0.4em\relax {IEEE}, 2019, pp. 1--20.
  [Online]. Available: \url{https://doi.org/10.1109/HOTCHIPS.2019.8875643}
\BIBentrySTDinterwordspacing

\bibitem{BannonVST19self}
\BIBentryALTinterwordspacing
P.~Bannon, G.~Venkataramanan, D.~D. Sarma, and E.~Talpes, ``Computer and
  redundancy solution for the full self-driving computer,'' in \emph{2019
  {IEEE} Hot Chips 31 Symposium (HCS), Cupertino, CA, USA, August 18-20,
  2019}.\hskip 1em plus 0.5em minus 0.4em\relax {IEEE}, 2019, pp. 1--22.
  [Online]. Available: \url{https://doi.org/10.1109/HOTCHIPS.2019.8875645}
\BIBentrySTDinterwordspacing

\bibitem{Huang2022}
\BIBentryALTinterwordspacing
C.~Huang, V.~J. Sorger, M.~Miscuglio, M.~Al-Qadasi, A.~Mukherjee, L.~Lampe,
  M.~Nichols, A.~N. Tait, T.~F.~D. Lima, B.~A. Marquez, J.~Wang,
  L.~Chrostowski, M.~P. Fok, D.~Brunner, S.~Fan, S.~Shekhar, P.~R. Prucnal, and
  B.~J. Shastri, ``Prospects and applications of photonic neural networks,''
  2022. [Online]. Available:
  \url{https://www.tandfonline.com/action/journalInformation?journalCode=tapx20}
\BIBentrySTDinterwordspacing

\bibitem{Tait2022quantify}
A.~N. Tait, ``Quantifying power in silicon photonic neural networks,''
  \emph{Physical Review Applied}, vol.~17, p. 054029, 5 2022.

\bibitem{hochberg2010towards}
M.~Hochberg and T.~Baehr-Jones, ``Towards fabless silicon photonics,''
  \emph{Nature photonics}, vol.~4, no.~8, pp. 492--494, 2010.

\bibitem{rahim2021taking}
A.~Rahim, A.~Hermans, B.~Wohlfeil, D.~Petousi, B.~Kuyken, D.~Van~Thourhout, and
  R.~G. Baets, ``Taking silicon photonics modulators to a higher performance
  level: state-of-the-art and a review of new technologies,'' \emph{Advanced
  Photonics}, vol.~3, no.~2, p. 024003, 2021.

\bibitem{Lischke2022}
\BIBentryALTinterwordspacing
S.~Lischke, A.~Peczek, J.~S. Morgan, K.~Sun, D.~Steckler, Y.~Yamamoto,
  F.~Korndamp, C.~Mai, S.~Marschmeyer, M.~Fraschke, A.~Kramp, A.~Beling, and
  L.~Zimmermann, ``Ultra-fast germanium photodiode with 3-db bandwidth of 265
  ghz.'' [Online]. Available: \url{https://doi.org/10.1038/s41566-021-00893-w}
\BIBentrySTDinterwordspacing

\bibitem{miscuglio2020massively}
M.~Miscuglio, Z.~Hu, S.~Li, J.~K. George, R.~Capanna, H.~Dalir, P.~M. Bardet,
  P.~Gupta, and V.~J. Sorger, ``Massively parallel amplitude-only fourier
  neural network,'' \emph{Optica}, vol.~7, no.~12, pp. 1812--1819, 2020.

\bibitem{peserico2022design}
N.~Peserico, H.~Yang, X.~Ma, S.~Li, M.~Hosseini, J.~K. George, P.~Gupta, C.~W.
  Wong, and V.~J. Sorger, ``Design and testing of integrated photonic chip for
  convolution neural network,'' in \emph{Imaging Systems and
  Applications}.\hskip 1em plus 0.5em minus 0.4em\relax Optica Publishing
  Group, 2022, pp. ITh3D--7.

\bibitem{cheng2020silicon}
Q.~Cheng, J.~Kwon, M.~Glick, M.~Bahadori, L.~P. Carloni, and K.~Bergman,
  ``Silicon photonics codesign for deep learning,'' \emph{Proceedings of the
  IEEE}, vol. 108, no.~8, pp. 1261--1282, 2020.

\bibitem{morichetti2014breakthroughs}
F.~Morichetti, S.~Grillanda, and A.~Melloni, ``Breakthroughs in photonics 2013:
  toward feedback-controlled integrated photonics,'' \emph{IEEE Photonics
  Journal}, vol.~6, no.~2, pp. 1--6, 2014.

\bibitem{zhou2022photonic}
H.~Zhou, J.~Dong, J.~Cheng, W.~Dong, C.~Huang, Y.~Shen, Q.~Zhang, M.~Gu,
  C.~Qian, H.~Chen \emph{et~al.}, ``Photonic matrix multiplication lights up
  photonic accelerator and beyond,'' \emph{Light: Science \& Applications},
  vol.~11, no.~1, pp. 1--21, 2022.

\bibitem{moss2022photonic}
D.~Moss, ``Photonic multiplexing architectures for optical neuromorphic
  computation,'' 2022.

\bibitem{Alqadasi2022}
\BIBentryALTinterwordspacing
M.~A. Al-Qadasi, L.~Chrostowski, and B.~J. Shastri, ``Scaling up silicon
  photonic-based accelerators: Challenges and opportunities collections
  articles you may be interested in,'' \emph{APL Photonics}, vol.~7, p. 20902,
  2022. [Online]. Available: \url{https://doi.org/10.1063/5.0070992}
\BIBentrySTDinterwordspacing

\bibitem{Zhou2022}
\BIBentryALTinterwordspacing
H.~Zhou, J.~Dong, J.~Cheng, W.~Dong, C.~Huang, Y.~Shen, Q.~Zhang, M.~Gu,
  C.~Qian, H.~Chen, Z.~Ruan, and X.~Zhang, ``Photonic matrix multiplication
  lights up photonic accelerator and beyond,'' \emph{Light: Science \&
  Applications 2022 11:1}, vol.~11, pp. 1--21, 2 2022. [Online]. Available:
  \url{https://www.nature.com/articles/s41377-022-00717-8}
\BIBentrySTDinterwordspacing

\bibitem{margalit2021perspective}
N.~Margalit, C.~Xiang, S.~M. Bowers, A.~Bjorlin, R.~Blum, and J.~E. Bowers,
  ``Perspective on the future of silicon photonics and electronics,''
  \emph{Applied Physics Letters}, vol. 118, no.~22, p. 220501, 2021.

\bibitem{Bogaerts2018}
W.~Bogaerts and L.~Chrostowski, ``Silicon photonics circuit design: Methods,
  tools and challenges,'' \emph{Laser \& Photonics Reviews}, vol.~12, p.
  1700237, 4 2018.

\bibitem{Chrostowski2019}
L.~Chrostowski, H.~Shoman, M.~Hammood, H.~Yun, J.~Jhoja, E.~Luan, S.~Lin,
  A.~Mistry, D.~Witt, N.~A.~F. Jaeger, S.~Shekhar, H.~Jayatilleka, P.~Jean,
  S.~B. de~Villers, J.~Cauchon, W.~Shi, C.~Horvath, J.~N. Westwood-Bachman,
  K.~Setzer, M.~Aktary, N.~S. Patrick, R.~J. Bojko, A.~Khavasi, X.~Wang, T.~F.
  de~Lima, A.~N. Tait, P.~R. Prucnal, D.~E. Hagan, D.~Stevanovic, and A.~P.
  Knights, ``Silicon photonic circuit design using rapid prototyping foundry
  process design kits,'' \emph{IEEE Journal of Selected Topics in Quantum
  Electronics}, vol.~25, pp. 1--26, 5 2019.

\bibitem{rios2015integrated}
C.~R{\'\i}os, M.~Stegmaier, P.~Hosseini, D.~Wang, T.~Scherer, C.~D. Wright,
  H.~Bhaskaran, and W.~H. Pernice, ``Integrated all-photonic non-volatile
  multi-level memory,'' \emph{Nature photonics}, vol.~9, no.~11, pp. 725--732,
  2015.

\bibitem{cheng2018device}
Z.~Cheng, C.~R{\'\i}os, N.~Youngblood, C.~D. Wright, W.~H. Pernice, and
  H.~Bhaskaran, ``Device-level photonic memories and logic applications using
  phase-change materials,'' \emph{Advanced Materials}, vol.~30, no.~32, p.
  1802435, 2018.

\bibitem{Reck1994}
\BIBentryALTinterwordspacing
M.~Reck, A.~Zeilinger, H.~J. Bernstein, and P.~Bertani, ``Experimental
  realization of any discrete unitary operator,'' \emph{Physical Review
  Letters}, vol.~73, p.~58, 7 1994. [Online]. Available:
  \url{https://journals.aps.org/prl/abstract/10.1103/PhysRevLett.73.58}
\BIBentrySTDinterwordspacing

\bibitem{miller2013self}
D.~A. Miller, ``Self-configuring universal linear optical component,''
  \emph{Photonics Research}, vol.~1, no.~1, pp. 1--15, 2013.

\bibitem{Metcalf2016}
B.~J. Metcalf, I.~A. Walmsley, P.~C. Humphreys, W.~S. Kolthammer, and W.~R.
  Clements, ``Optimal design for universal multiport interferometers,''
  \emph{Optica, Vol. 3, Issue 12, pp. 1460-1465}, vol.~3, pp. 1460--1465, 12
  2016.

\bibitem{Shen2017}
Y.~Shen, N.~C. Harris, S.~Skirlo, M.~Prabhu, T.~Baehr-Jones, M.~Hochberg,
  X.~Sun, S.~Zhao, H.~Larochelle, D.~Englund, and M.~Soljacic, ``Deep learning
  with coherent nanophotonic circuits,'' \emph{Nature Photonics}, vol.~11, pp.
  441--446, 6 2017.

\bibitem{Demirkiran2021}
\BIBentryALTinterwordspacing
C.~Demirkiran, F.~Eris, G.~Wang, J.~Elmhurst, N.~Moore, N.~C. Harris,
  A.~Basumallik, V.~J. Reddi, A.~Joshi, and D.~Bunandar, ``An electro-photonic
  system for accelerating deep neural networks,'' 9 2021. [Online]. Available:
  \url{http://arxiv.org/abs/2109.01126}
\BIBentrySTDinterwordspacing

\bibitem{Zhang2021}
\BIBentryALTinterwordspacing
H.~Zhang, M.~Gu, X.~D. Jiang, J.~Thompson, H.~Cai, S.~Paesani, R.~Santagati,
  A.~Laing, Y.~Zhang, M.~H. Yung, Y.~Z. Shi, F.~K. Muhammad, G.~Q. Lo, X.~S.
  Luo, B.~Dong, D.~L. Kwong, L.~C. Kwek, and A.~Q. Liu, ``An optical neural
  chip for implementing complex-valued neural network.'' [Online]. Available:
  \url{https://doi.org/10.1038/s41467-020-20719-7}
\BIBentrySTDinterwordspacing

\bibitem{Feng2021}
C.~Feng, J.~Gu, H.~Zhu, Z.~Ying, Z.~Zhao, D.~Z. Pan, and R.~T. Chen, ``A
  compact butterfly-style silicon photonic-electronic neural chip for
  hardware-efficient deep learning,'' 2021.

\bibitem{Bandyopadhyay2022}
S.~Bandyopadhyay, A.~Sludds, S.~Krastanov, R.~Hamerly, N.~Harris, D.~Bunandar,
  M.~Streshinsky, M.~Hochberg, and D.~Englund, ``Single chip photonic deep
  neural network with accelerated training,'' 2022.

\bibitem{Carolan2015}
J.~Carolan, C.~Harrold, C.~Sparrow, E.~Martín-López, N.~J. Russell, J.~W.
  Silverstone, P.~J. Shadbolt, N.~Matsuda, M.~Oguma, M.~Itoh, G.~D. Marshall,
  M.~G. Thompson, J.~C. Matthews, T.~Hashimoto, J.~L. O'Brien, and A.~Laing,
  ``Universal linear optics,'' \emph{Science}, vol. 349, pp. 711--716, 8 2015.

\bibitem{Ruocco2016}
A.~Ruocco, A.~Ribeiro, L.~Vanacker, and W.~Bogaerts, ``Demonstration of a
  4x4-port universal linear circuit,'' \emph{Optica, Vol. 3, Issue 12, pp.
  1348-1357}, vol.~3, pp. 1348--1357, 12 2016.

\bibitem{annoni2017unscrambling}
A.~Annoni, E.~Guglielmi, M.~Carminati, G.~Ferrari, M.~Sampietro, D.~A. Miller,
  A.~Melloni, and F.~Morichetti, ``Unscrambling light—automatically undoing
  strong mixing between modes,'' \emph{Light: Science \& Applications}, vol.~6,
  no.~12, pp. e17\,110--e17\,110, 2017.

\bibitem{miller2015perfect}
D.~A. Miller, ``Perfect optics with imperfect components,'' \emph{Optica},
  vol.~2, no.~8, pp. 747--750, 2015.

\bibitem{cem2022data}
A.~Cem, S.~Yan, Y.~Ding, D.~Zibar, and F.~Da~Ros, ``Data-driven modeling of
  mach-zehnder interferometer-based optical matrix multipliers,'' \emph{arXiv
  preprint arXiv:2210.09171}, 2022.

\bibitem{9252466}
F.~Shokraneh, S.~Geoffroy-Gagnon, and O.~Liboiron-Ladouceur, ``Towards
  phase-error- and loss-tolerant programmable mzi-based optical processors for
  optical neural networks,'' in \emph{2020 IEEE Photonics Conference (IPC)},
  2020, pp. 1--2.

\bibitem{9489417}
R.~Hamerly, S.~Bandyopadhyay, and D.~Englund, ``Robust zero-change
  self-configuration of the rectangular mesh,'' in \emph{2021 Optical Fiber
  Communications Conference and Exhibition (OFC)}, 2021, pp. 1--3.

\bibitem{Ji2012}
R.~Ji, J.~Ding, L.~Yang, L.~Zhang, and Q.~Xu, ``On-chip cmos-compatible optical
  signal processor,'' \emph{Optics Express, Vol. 20, Issue 12, pp.
  13560-13565}, vol.~20, pp. 13\,560--13\,565, 6 2012.

\bibitem{Feldmann2021}
\BIBentryALTinterwordspacing
J.~Feldmann, N.~Youngblood, M.~Karpov, H.~Gehring, X.~Li, M.~Stappers, M.~L.
  Gallo, X.~Fu, A.~Lukashchuk, A.~S. Raja, J.~Liu, C.~D. Wright, A.~Sebastian,
  T.~J. Kippenberg, W.~H.~P. Pernice, and H.~Bhaskaran, ``Parallel
  convolutional processing using an integrated photonic tensor core,''
  \emph{Nature}, vol. 589, 2021. [Online]. Available:
  \url{https://doi.org/10.1038/s41586-020-03070-1}
\BIBentrySTDinterwordspacing

\bibitem{Tait2014}
A.~N. Tait, M.~A. Nahmias, B.~J. Shastri, and P.~R. Prucnal, ``Broadcast and
  weight: An integrated network for scalable photonic spike processing,''
  \emph{Journal of Lightwave Technology}, vol.~32, pp. 4029--4041, 11 2014.

\bibitem{Liu2021}
\BIBentryALTinterwordspacing
G.~Liu, W.-P. Ma, H.~Cao, al, R.~Zhu, T.~Qiu, J.~Wang, B.~A. Marquez, Z.~Guo,
  H.~Morison, S.~Shekhar, L.~Chrostowski, P.~Prucnal, and B.~J. Shastri,
  ``Photonic pattern reconstruction enabled by on-chip online learning and
  inference,'' \emph{Journal of Physics: Photonics}, vol.~3, p. 024006, 2 2021.
  [Online]. Available:
  \url{https://iopscience.iop.org/article/10.1088/2515-7647/abe3d9
  https://iopscience.iop.org/article/10.1088/2515-7647/abe3d9/meta}
\BIBentrySTDinterwordspacing

\bibitem{ma2022high}
X.~Ma, N.~Peserico, A.~Khaled, Z.~Guo, B.~Nouri, H.~Dalir, B.~Shastri, and
  V.~Sorger, ``High-density integrated photonic tensor processing unit with a
  matrix multiply compiler,'' 2022.

\bibitem{Bruckerhoff2022}
\BIBentryALTinterwordspacing
F.~Brückerhoff-Plückelmann, J.~Feldmann, H.~Gehring, W.~Zhou, C.~D. Wright,
  H.~Bhaskaran, and W.~Pernice, ``Broadband photonic tensor core with
  integrated ultra-low crosstalk wavelength multiplexers,''
  \emph{Nanophotonics}, vol.~0, 2 2022. [Online]. Available:
  \url{https://www.degruyter.com/document/doi/10.1515/nanoph-2021-0752/html}
\BIBentrySTDinterwordspacing

\bibitem{Sarwat2022}
\BIBentryALTinterwordspacing
S.~G. Sarwat, F.~Brückerhoff-Plückelmann, S.~G.~C. Carrillo, E.~Gemo,
  J.~Feldmann, H.~Bhaskaran, C.~D. Wright, W.~H. Pernice, and A.~Sebastian,
  ``An integrated photonics engine for unsupervised correlation detection,''
  \emph{Science Advances}, vol.~8, p. 3243, 6 2022. [Online]. Available:
  \url{https://www.science.org/doi/10.1126/sciadv.abn3243}
\BIBentrySTDinterwordspacing

\bibitem{Xu_11}
\BIBentryALTinterwordspacing
Q.~Xu and R.~Soref, ``Reconfigurable optical directed-logic circuits using
  microresonator-based optical switches,'' \emph{Opt. Express}, vol.~19, no.~6,
  pp. 5244--5259, Mar 2011. [Online]. Available:
  \url{https://opg.optica.org/oe/abstract.cfm?URI=oe-19-6-5244}
\BIBentrySTDinterwordspacing

\bibitem{Ibrahim2004}
T.~A. Ibrahim, K.~Amarnath, L.~C. Kuo, R.~Grover, V.~Van, and P.-T. Ho,
  ``Photonic logic nor gate based on two symmetric microring resonators,''
  \emph{Optics Letters}, vol.~29, p. 2779, 12 2004.

\bibitem{Tait2017}
\BIBentryALTinterwordspacing
A.~N. Tait, T.~F.~D. Lima, E.~Zhou, A.~X. Wu, M.~A. Nahmias, B.~J. Shastri, and
  P.~R. Prucnal, ``Neuromorphic photonic networks using silicon photonic weight
  banks,'' \emph{Scientific Reports 2017 7:1}, vol.~7, pp. 1--10, 8 2017.
  [Online]. Available: \url{https://www.nature.com/articles/s41598-017-07754-z}
\BIBentrySTDinterwordspacing

\bibitem{huang2021silicon}
C.~Huang, S.~Fujisawa, T.~F. de~Lima, A.~N. Tait, E.~C. Blow, Y.~Tian,
  S.~Bilodeau, A.~Jha, F.~Yaman, H.-T. Peng \emph{et~al.}, ``A silicon
  photonic--electronic neural network for fibre nonlinearity compensation,''
  \emph{Nature Electronics}, vol.~4, no.~11, pp. 837--844, 2021.

\bibitem{Zhang2022}
W.~Zhang, C.~Huang, C.~Huang, H.-T. Peng, S.~Bilodeau, A.~Jha, E.~Blow, T.~F.
  de~Lima, T.~F. de~Lima, B.~J. Shastri, B.~J. Shastri, and P.~Prucnal,
  ``Silicon microring synapses enable photonic deep learning beyond 9-bit
  precision,'' \emph{Optica, Vol. 9, Issue 5, pp. 579-584}, vol.~9, pp.
  579--584, 5 2022.

\bibitem{amin2018waveguide}
R.~Amin, J.~B. Khurgin, and V.~J. Sorger, ``Waveguide-based electro-absorption
  modulator performance: comparative analysis,'' \emph{Optics express},
  vol.~26, no.~12, pp. 15\,445--15\,470, 2018.

\bibitem{alexoudi2020optical}
T.~Alexoudi, G.~T. Kanellos, and N.~Pleros, ``Optical ram and integrated
  optical memories: a survey,'' \emph{Light: Science \& Applications}, vol.~9,
  no.~1, pp. 1--16, 2020.

\bibitem{Peserico2022}
N.~Peserico, T.~F. de~Lima, T.~F. de~Lima, P.~Prucnal, and V.~J. Sorger,
  ``Emerging devices and packaging strategies for electronic-photonic ai
  accelerators: opinion,'' \emph{Optical Materials Express, Vol. 12, Issue 4,
  pp. 1347-1351}, vol.~12, pp. 1347--1351, 4 2022.

\bibitem{meng2022electrical}
J.~Meng, M.~Miscuglio, N.~Peserico, X.~Ma, Y.~Zhang, C.-C. Popescu, M.~Kang,
  K.~Richardson, J.~Hu, and V.~J. Sorger, ``Electrical pulse driven multi-level
  nonvolatile photonic memories using broadband transparent phase change
  materials,'' \emph{arXiv preprint arXiv:2203.13337}, 2022.

\bibitem{sahoo2022gsst}
D.~Sahoo and R.~Naik, ``Gsst phase change materials and its utilization in
  optoelectronic devices: A review,'' \emph{Materials Research Bulletin}, vol.
  148, p. 111679, 2022.

\bibitem{redaelli2022material}
A.~Redaelli, E.~Petroni, and R.~Annunziata, ``Material and process engineering
  challenges in ge-rich gst for embedded pcm,'' \emph{Materials Science in
  Semiconductor Processing}, vol. 137, p. 106184, 2022.

\bibitem{Ashtiani2022}
\BIBentryALTinterwordspacing
F.~Ashtiani, A.~J. Geers, and F.~Aflatouni, ``An on-chip photonic deep neural
  network for image classification,'' \emph{Nature}, vol. 606, 2022. [Online].
  Available: \url{https://doi.org/10.1038/s41586-022-04714-0}
\BIBentrySTDinterwordspacing

\bibitem{tait2019silicon}
A.~N. Tait, T.~F. De~Lima, M.~A. Nahmias, H.~B. Miller, H.-T. Peng, B.~J.
  Shastri, and P.~R. Prucnal, ``Silicon photonic modulator neuron,''
  \emph{Physical Review Applied}, vol.~11, no.~6, p. 064043, 2019.

\bibitem{Moayedi2020}
M.~Moayedi, P.~Fard, I.~A.~D. Williamson, M.~Edwards, K.~E. Liu, S.~Pai,
  B.~Bartlett, M.~Minkov, T.~W. Hughes, S.~Fan, and T.-A. Nguyen,
  ``Experimental realization of arbitrary activation functions for optical
  neural networks,'' \emph{Optics Express, Vol. 28, Issue 8, pp. 12138-12148},
  vol.~28, pp. 12\,138--12\,148, 4 2020.

\bibitem{Shi2022}
\BIBentryALTinterwordspacing
Y.~Shi, J.~Ren, G.~Chen, W.~Liu, C.~Jin, X.~Guo, Y.~Yu, and X.~Zhang,
  ``Nonlinear germanium-silicon photodiode for activation and monitoring in
  photonic neuromorphic networks.'' [Online]. Available:
  \url{https://doi.org/10.1038/s41467-022-33877-7}
\BIBentrySTDinterwordspacing

\bibitem{gupta20224f}
P.~Gupta and S.~Li, ``4f optical neural network acceleration: an architecture
  perspective,'' in \emph{AI and Optical Data Sciences III}, vol. 12019.\hskip
  1em plus 0.5em minus 0.4em\relax SPIE, 2022, pp. 77--84.

\bibitem{leuthold2010nonlinear}
J.~Leuthold, C.~Koos, and W.~Freude, ``Nonlinear silicon photonics,''
  \emph{Nature photonics}, vol.~4, no.~8, pp. 535--544, 2010.

\bibitem{morichetti2011travelling}
F.~Morichetti, A.~Canciamilla, C.~Ferrari, A.~Samarelli, M.~Sorel, and
  A.~Melloni, ``Travelling-wave resonant four-wave mixing breaks the limits of
  cavity-enhanced all-optical wavelength conversion,'' \emph{Nature
  communications}, vol.~2, no.~1, pp. 1--8, 2011.

\bibitem{chang2022integrated}
L.~Chang, S.~Liu, and J.~E. Bowers, ``Integrated optical frequency comb
  technologies,'' \emph{Nature Photonics}, vol.~16, no.~2, pp. 95--108, 2022.

\bibitem{Thomas2020Noise}
T.~F. de~Lima, A.~N. Tait, H.~Saeidi, M.~A. Nahmias, H.~T. Peng, S.~Abbaslou,
  B.~J. Shastri, and P.~R. Prucnal, ``Noise analysis of photonic modulator
  neurons,'' \emph{IEEE Journal of Selected Topics in Quantum Electronics},
  vol.~26, 1 2020.

\bibitem{amin2019ito}
R.~Amin, J.~George, S.~Sun, T.~Ferreira~de Lima, A.~N. Tait, J.~Khurgin,
  M.~Miscuglio, B.~J. Shastri, P.~R. Prucnal, T.~El-Ghazawi \emph{et~al.},
  ``Ito-based electro-absorption modulator for photonic neural activation
  function,'' \emph{APL Materials}, vol.~7, no.~8, p. 081112, 2019.

\bibitem{gui2022100}
Y.~Gui, B.~M. Nouri, M.~Miscuglio, R.~Amin, H.~Wang, J.~B. Khurgin, H.~Dalir,
  and V.~J. Sorger, ``100 ghz micrometer-compact broadband monolithic ito
  mach--zehnder interferometer modulator enabling 3500 times higher packing
  density,'' \emph{Nanophotonics}, 2022.

\bibitem{amin2021ito}
R.~Amin, J.~K. George, H.~Wang, R.~Maiti, Z.~Ma, H.~Dalir, J.~B. Khurgin, and
  V.~J. Sorger, ``An ito--graphene heterojunction integrated absorption
  modulator on si-photonics for neuromorphic nonlinear activation,'' \emph{APL
  Photonics}, vol.~6, no.~12, p. 120801, 2021.

\bibitem{feldmann2019all}
J.~Feldmann, N.~Youngblood, C.~D. Wright, H.~Bhaskaran, and W.~H. Pernice,
  ``All-optical spiking neurosynaptic networks with self-learning
  capabilities,'' \emph{Nature}, vol. 569, no. 7755, pp. 208--214, 2019.

\bibitem{Xu2022}
\BIBentryALTinterwordspacing
Z.~Xu, B.~Tang, X.~Zhang, J.~F. Leong, J.~Pan, S.~Hooda, E.~Zamburg, and
  A.~V.-Y. Thean, ``Reconfigurable nonlinear photonic activation function for
  photonic neural network based on non-volatile opto-resistive ram switch,''
  \emph{Official journal of the CIOMP}, pp. 2047--7538. [Online]. Available:
  \url{www.nature.com/lsa}
\BIBentrySTDinterwordspacing

\bibitem{lischke2021ultra}
S.~Lischke, A.~Peczek, J.~Morgan, K.~Sun, D.~Steckler, Y.~Yamamoto,
  F.~Kornd{\"o}rfer, C.~Mai, S.~Marschmeyer, M.~Fraschke \emph{et~al.},
  ``Ultra-fast germanium photodiode with 3-db bandwidth of 265 ghz,''
  \emph{Nature Photonics}, vol.~15, no.~12, pp. 925--931, 2021.

\bibitem{fratalocchi2015nano}
A.~Fratalocchi, C.~M. Dodson, R.~Zia, P.~Genevet, E.~Verhagen, H.~Altug, and
  V.~J. Sorger, ``Nano-optics gets practical,'' \emph{Nature Nanotechnology},
  vol.~10, no. ARTICLE, pp. 11--15, 2015.

\bibitem{wang2022high}
H.~Wang, M.~Thomaschewski, C.~Patil, Y.~Gui, B.~M. Nouri, H.~Dalir, and V.~J.
  Sorger, ``High-performance opto-electronics with emerging materials,'' in
  \emph{Low-Dimensional Materials and Devices 2022}, vol. 12200.\hskip 1em plus
  0.5em minus 0.4em\relax SPIE, 2022, p. 1220002.

\bibitem{amin2020sub}
R.~Amin, R.~Maiti, Y.~Gui, C.~Suer, M.~Miscuglio, E.~Heidari, R.~T. Chen,
  H.~Dalir, and V.~J. Sorger, ``Sub-wavelength ghz-fast broadband ito
  mach--zehnder modulator on silicon photonics,'' \emph{Optica}, vol.~7, no.~4,
  pp. 333--335, 2020.

\bibitem{pickus2013silicon}
S.~K. Pickus, S.~Khan, C.~Ye, Z.~Li, and V.~J. Sorger, ``Silicon plasmon
  modulators: breaking photonic limits.''

\bibitem{sorger2020roadmap}
V.~J. Sorger and R.~Maiti, ``Roadmap for gain-bandwidth-product enhanced
  photodetectors: opinion,'' \emph{Optical Materials Express}, vol.~10, no.~9,
  pp. 2192--2200, 2020.

\bibitem{patil2022highly}
C.~Patil, H.~Dalir, J.~H. Kang, A.~Davydov, C.~W. Wong, and V.~J. Sorger,
  ``Highly accurate, reliable, and non-contaminating two-dimensional material
  transfer system,'' \emph{Applied Physics Reviews}, vol.~9, no.~1, p. 011419,
  2022.

\bibitem{wang2022self}
H.~Wang, Y.~Gui, C.~Dong, S.~Altaleb, B.~M. Nouri, M.~Thomaschewski, H.~Dalir,
  and V.~J. Sorger, ``Self-powered broadband photodetector based on mos2/sb2te3
  heterojunctions: a promising approach for highly sensitive detection,''
  \emph{Nanophotonics}, 2022.

\bibitem{wang2017novel}
Z.~Wang, A.~Abbasi, U.~Dave, A.~De~Groote, S.~Kumari, B.~Kunert, C.~Merckling,
  M.~Pantouvaki, Y.~Shi, B.~Tian \emph{et~al.}, ``Novel light source
  integration approaches for silicon photonics,'' \emph{Laser \& Photonics
  Reviews}, vol.~11, no.~4, p. 1700063, 2017.

\bibitem{meng2021electronic}
J.~Meng, M.~Miscuglio, J.~K. George, A.~Babakhani, and V.~J. Sorger,
  ``Electronic bottleneck suppression in next-generation networks with
  integrated photonic digital-to-analog converters,'' \emph{Advanced photonics
  research}, vol.~2, no.~2, p. 2000033, 2021.

\bibitem{jalali2006raman}
B.~Jalali, V.~Raghunathan, D.~Dimitropoulos, and O.~Boyraz, ``Raman-based
  silicon photonics,'' \emph{IEEE Journal of Selected Topics in Quantum
  Electronics}, vol.~12, no.~3, pp. 412--421, 2006.

\bibitem{tahersima2015enhanced}
M.~H. Tahersima and V.~J. Sorger, ``Enhanced photon absorption in spiral
  nanostructured solar cells using layered 2d materials,''
  \emph{Nanotechnology}, vol.~26, no.~34, p. 344005, 2015.

\bibitem{narayana2017morphonoc}
V.~K. Narayana, S.~Sun, A.-H.~A. Badawy, V.~J. Sorger, and T.~El-Ghazawi,
  ``Morphonoc: Exploring the design space of a configurable hybrid noc using
  nanophotonics,'' \emph{Microprocessors and Microsystems}, vol.~50, pp.
  113--126, 2017.

\bibitem{shen2022reconfigurable}
C.~Shen, N.~Peserico, J.~Meng, X.~Ma, B.~M. Nouri, C.-C. Popescu, J.~Hu,
  T.~El-Ghazawi, H.~Dalir, and V.~J. Sorger, ``Reconfigurable
  application-specific photonic integrated circuit for solving partial
  differential equations,'' \emph{arXiv preprint arXiv:2208.03588}, 2022.

\bibitem{9083571}
M.~Rakowski, C.~Meagher, K.~Nummy, A.~Aboketaf, J.~Ayala, Y.~Bian, B.~Harris,
  K.~Mclean, K.~McStay, A.~Sahin, L.~Medina, B.~Peng, Z.~Sowinski, A.~Stricker,
  T.~Houghton, C.~Hedges, K.~Giewont, A.~Jacob, T.~Letavic, D.~Riggs, A.~Yu,
  and J.~Pellerin, ``45nm cmos — silicon photonics monolithic technology
  (45clo) for next-generation, low power and high speed optical
  interconnects,'' in \emph{2020 Optical Fiber Communications Conference and
  Exhibition (OFC)}, 2020, pp. 1--3.

\bibitem{chan2016beol}
W.-T.~J. Chan, Y.~Du, A.~B. Kahng, S.~Nath, and K.~Samadi, ``Beol stack-aware
  routability prediction from placement using data mining techniques,'' in
  \emph{2016 IEEE 34th International Conference on Computer Design
  (ICCD)}.\hskip 1em plus 0.5em minus 0.4em\relax IEEE, 2016, pp. 41--48.

\bibitem{standaert2016beol}
T.~Standaert, G.~Beique, H.-C. Chen, S.-T. Chen, B.~Hamieh, J.~Lee,
  P.~McLaughlin, J.~McMahon, Y.~Mignot, F.~Mont \emph{et~al.}, ``Beol process
  integration for the 7 nm technology node,'' in \emph{2016 IEEE international
  interconnect technology conference/advanced metallization conference
  (IITC/AMC)}.\hskip 1em plus 0.5em minus 0.4em\relax IEEE, 2016, pp. 2--4.

\bibitem{vega2019three}
V.~Vega-Gonzalez, C.~Wilson, B.~Briggs, S.~Decoster, J.~Versluijs,
  A.~Le{\'s}niewska, S.~Paolillo, R.~Baert, H.~Puliyalil, J.~Bekaert
  \emph{et~al.}, ``Three-layer beol process integration with supervia and
  self-aligned-block options for the 3 nm node,'' in \emph{2019 IEEE
  International Electron Devices Meeting (IEDM)}.\hskip 1em plus 0.5em minus
  0.4em\relax IEEE, 2019, pp. 19--3.

\bibitem{billah2018hybrid}
M.~R. Billah, M.~Blaicher, T.~Hoose, P.-I. Dietrich, P.~Marin-Palomo,
  N.~Lindenmann, A.~Nesic, A.~Hofmann, U.~Troppenz, M.~Moehrle \emph{et~al.},
  ``Hybrid integration of silicon photonics circuits and inp lasers by photonic
  wire bonding,'' \emph{Optica}, vol.~5, no.~7, pp. 876--883, 2018.

\end{thebibliography}

\begin{IEEEbiography}[{\includegraphics[width=1in,height=1.25in,clip,keepaspectratio]{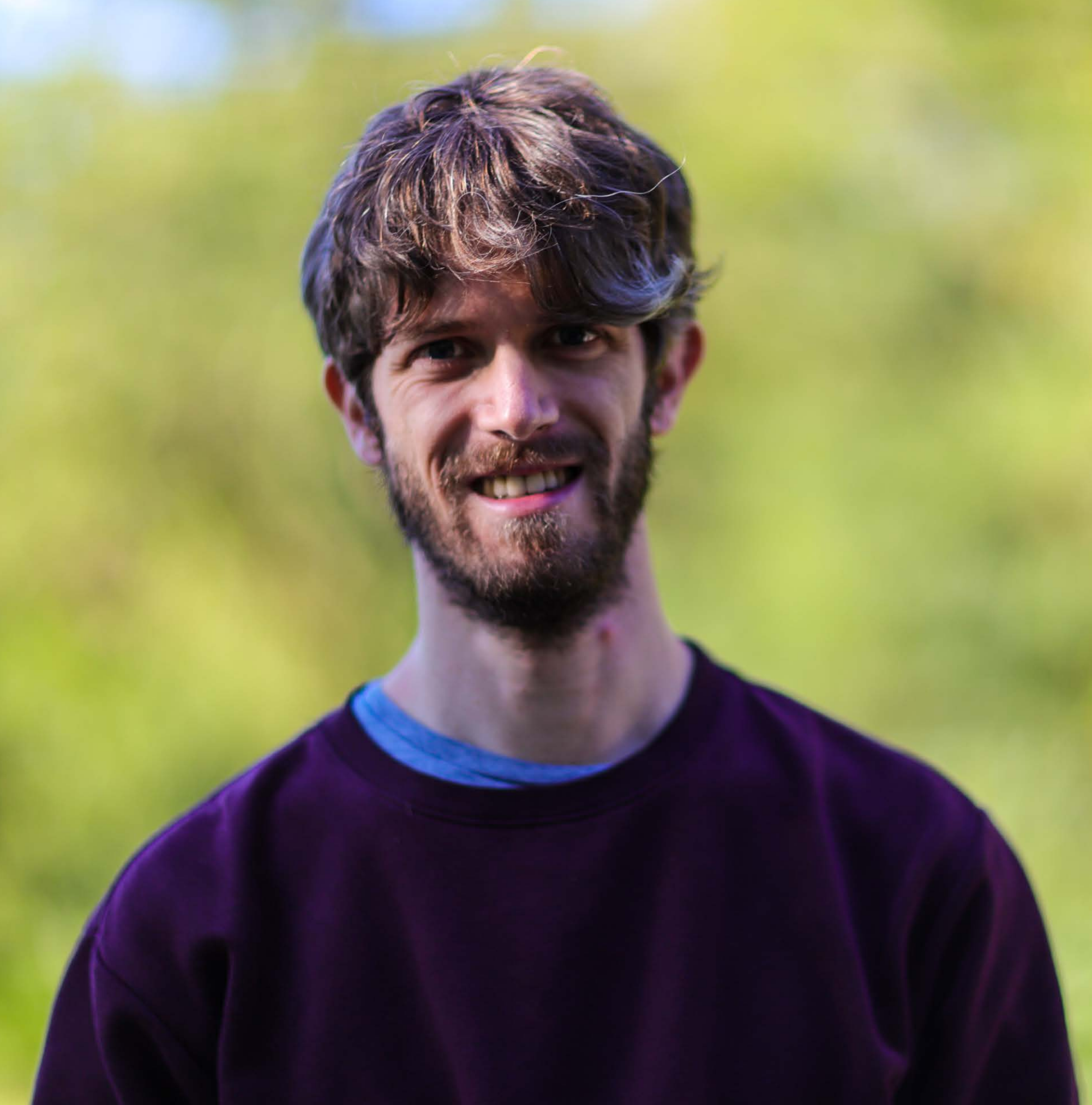}}]{Dr. Nicola Peserico}
received his PhD at Politecnico di Milano (Italy) in 2018. In 2019, he joined Femtorays (Italy), a silicon photonics startup for biosensing. He is now a Post-doc Researcher in the Department of Electrical and Computer Engineering at the George Washington University, Washington, DC. His research area include silicon photonics, AI/ML  accelerators, optoelectronics devices and components, and bio-sensing with photonic integrated circuits.
\end{IEEEbiography}



\begin{IEEEbiography}[{\includegraphics[width=1in,height=1.25in,clip,keepaspectratio]{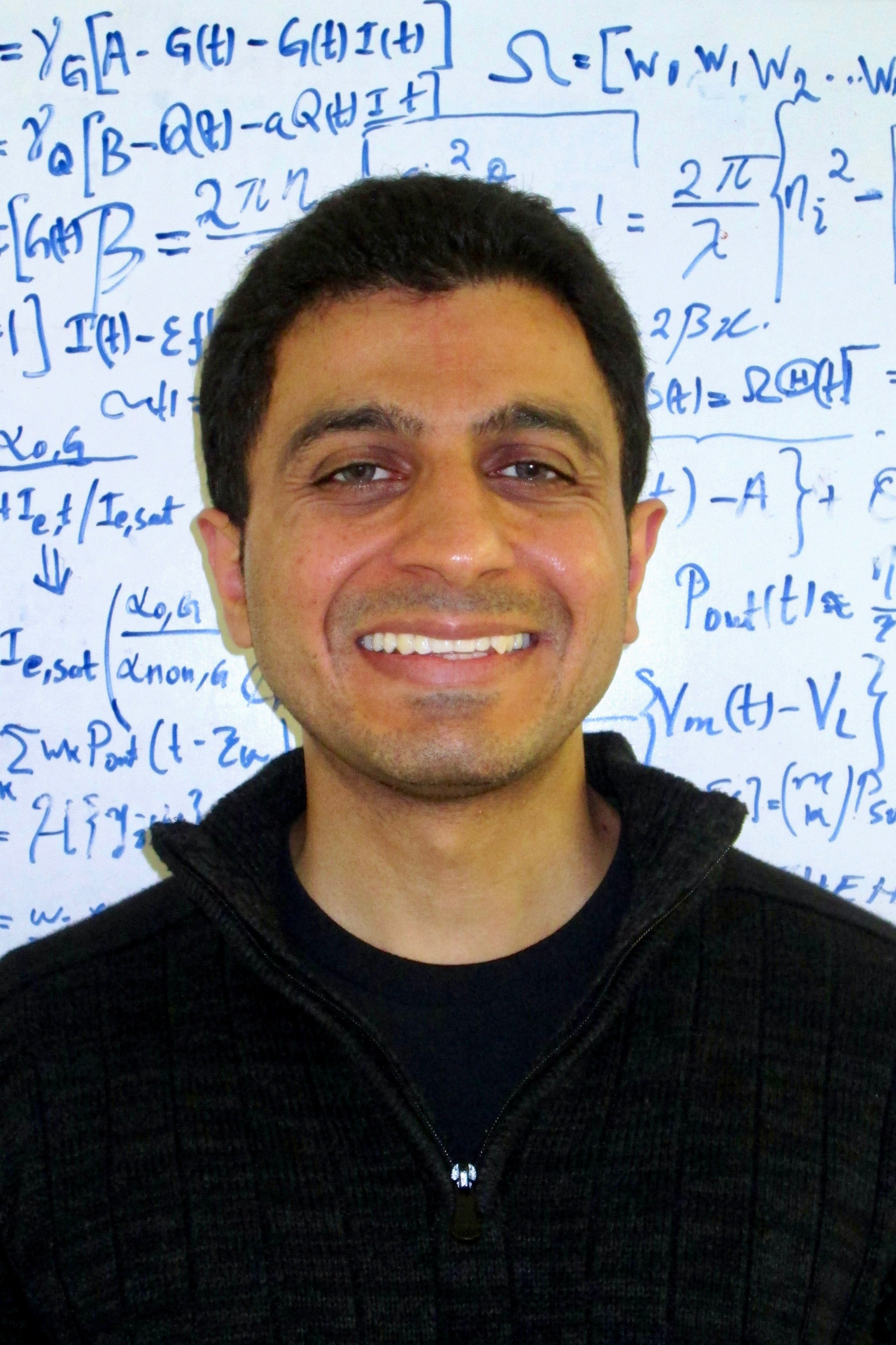}}]{Dr. Bhavin Shastri}
is an Assistant Professor of Engineering Physics at Queen's University, Canada, and a Faculty Affiliate at the Vector Institute for Artificial Intelligence, Canada. He was an Associate Research Scholar (2016-2018) and Banting/NSERC Postdoctoral Fellow (2012-2016) at Princeton University. He received a Ph.D. degree in electrical engineering (photonics) from McGill University in 2012. With research interests in silicon photonics, photonic integrated circuits, neuromorphic computing, and machine learning, he has published more than 70 journal articles and 100 conference proceedings, seven book chapters, and given over 65 invited talks and lectures, five keynotes and four tutorials. He is a co-author of the book (CRC Press, 2017) Neuromorphic Photonics, a term he helped coin. Dr. Shastri is the recipient of the 2022 SPIE Early Career Achievement Award and the 2020 IUPAP Young Scientist Prize in Optics "for his pioneering contributions to neuromorphic photonics" from ICO. He is a Senior Member of Optica (formerly OSA) and IEEE, recipient of the 2014 Banting Postdoctoral Fellowship from the Government of Canada, the 2012 D. W. Ambridge Prize for the top graduating Ph.D. student at McGill, an IEEE Photonics Society 2011 Graduate Student Fellowship amongst others awards.
\end{IEEEbiography}

\begin{IEEEbiography}[{\includegraphics[width=1in,height=1.25in,clip,keepaspectratio]{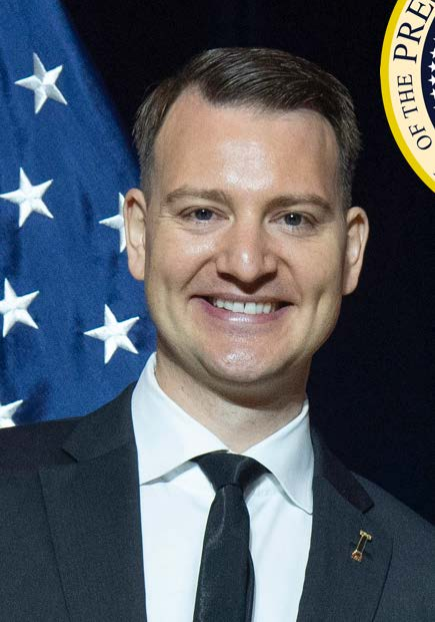}}]{Dr. Volker J. Sorger}
is a Full Professor in the Department of Electrical and Computer Engineering and the Director of the Institute on AI \& Photonics, the Head of the Devices  \& Intelligent Systems Laboratory at the George Washington University. His research areas include devices \& optoelectronics, AI/ML accelerators, mixed-signal ASICs, quantum matter \& quantum processors, cryptography. For his work, Dr. Sorger received multiple awards including the Presidential PECASE Award, the AFOSR YIP, the Emil Wolf Prize, and the National Academy of Sciences award of the year. Dr. Sorger is Editor for Optica, Nanophotonics, Applied Physics Rev., eLight, Chips, and was the former editor-in-chief of Nanophotonics. He is a Fellow of Optica (former OSA), a Fellow of SPIE, a Fellow of the German National Academic Foundation, and a Senior Member of IEEE. He is a founder of Optelligence.
\end{IEEEbiography}





\end{document}